# Distributed dynamic modeling and monitoring for large-scale industrial processes under closed-loop control

Wenqing Li[1], Chunhui Zhao[1,2*], Biao Huang[3]

1. *State Key Laboratory of Industrial Control Technology, College of Control Science and Engineering, Zhejiang University, Hangzhou, 310027, China*

2. *Hubei Key Laboratory of Advanced Control and Intelligent Automation for Complex Systems*

3. *Department of Chemical and Materials Engineering, University of Alberta, Edmonton, AB, T6G-2V4, Canada*

**Abstract**—For large-scale industrial processes under closed-loop control, process dynamics directly resulting from control action are typical characteristics and may show different behaviors between real faults and normal changes of operating conditions. However, conventional distributed monitoring approaches do not consider the closed-loop control mechanism and only explore static characteristics, which thus are incapable of distinguishing between real process faults and nominal changes of operating conditions, leading to unnecessary alarms. In this regard, this paper proposes a distributed monitoring method for closed-loop industrial processes by concurrently exploring static and dynamic characteristics. First, the large-scale closed-loop process is decomposed into several subsystems by developing a sparse slow feature analysis (SSFA) algorithm which capture changes of both static and dynamic information. Second, distributed models are developed to separately capture static and dynamic characteristics from the local and global aspects. Based on the distributed monitoring system, a two-level monitoring strategy is proposed to check different influences on process characteristics resulting from changes of the operating conditions and control action, and thus the two changes can be well distinguished from each other. Case studies are conducted based on both benchmark data and real industrial process data to illustrate the effectiveness of the proposed method.

**1 Introduction**

With the recent development of measurement devices and data technologies, data-based process monitoring[1-9] becomes quite popular in both industrial application and academic research. Compared with model-based monitoring approaches, data-based methods are less dependent on process knowledge which becomes increasingly difficult to obtain. If the critical information of the process is well captured by the data, the process could be well modeled and monitoring schemes could be well established. However, traditional data-based methods[10-13] may not function well for large-scale processes. The problem results from the fact that industrial data are collected from different parts of the whole process with different operation patterns, indicating different variable correlations. As a result, process characteristics cannot be captured by a centralized model. Therefore, distributed approaches[14-19] with multi-block decomposition have received great attention. Compared with the centralized methods, distributed approaches have several advantages. First, by dividing the whole process into different blocks, model accuracy may be improved since variables with similar characteristics are assembled in the same block. Second, the distributed framework provides an easier and more efficient way for fault detection because it can be readily implemented in each local block individually.

While the distributed methods have been widely used for large-scale processes, most of the works[14-16] rely on division of the sub-blocks based on limited prior process knowledge, which may be subject to physical constraints, process topologies, etc. Therefore, alternative approaches with automatic partition criterions[17-19] have been established. Tong et al.[17] derived four subspaces from the input spaces by evaluating the relevance between variables and principal components of PCA. Jiang et al.[18] used the fault-relevant relative analysis index[20] for variable selection, in which different variable subsets are determined by analyzing the relationships between variables and specific faults. Considering that hybrid correlations may exist (coexist of linearity and nonlinearity) among variables, Li et al.[19] proposed a linear variable subset partition algorithm, which achieved sub-block partition by decoupling the hybrid correlations into linear part and nonlinear part. Despite that the above distributed methods[17-19] are successful, they only analyze static characteristics and ignore process dynamics which reveal control action of closed-loop systems. For closed-loop processes, static distribution of process data reflects steady-state operating conditions, and it would change significantly when process deviates from its predefined conditions caused by either process faults or nominal changes of operating conditions. On the contrary, the feedback control mechanism may result in different dynamic variations between nominal operating condition changes and real process faults. More specifically, the disturbance caused by normal operating condition changes can be well compensated and the manipulated variables may be finally controlled around their new setpoints, resulting in approximately invariable dynamic variations. For real process faults, unusual dynamic variations may be observed since the disturbance is too serious to compensate. Therefore, both the static and dynamic characteristics should be explored to reveal the effect of control mechanism for closed-loop industrial processes.

To capture process dynamics, dynamic PCA[21], dynamic PLS[22] and state-space based approaches[23,24] have been proposed, which show superior performance to classic multivariate statistical methods. Nevertheless, these methods mix static and dynamic information and lack explicit descriptions of process dynamics, which could jeopardize monitoring performance as critical information might be buried. Considering that static information and dynamic information reveal different aspects of closed-loop processes, a slow feature analysis (SFA) based monitoring strategy[25] was proposed by Shang et al., where slow features are developed with clear temporal interpretations based on which two groups of indices are calculated for separately monitoring static and dynamic characteristics. As a result, SFA based monitoring schemes[25,26] can well distinguish normal operating condition changes from real faults for closed-loop industrial processes. However, similar to PCA, they may not work well when dealing with large-scale closed-loop processes.

For large-scale closed-loop industrial processes, variables belonging to the same subsystem may have strong static and dynamic correlations thanks to the effects of control action. Here, static and dynamic correlations denotes the correlations among measured variables and the temporary correlations among measured variables respectively. Therefore, there are two critical problems that should be considered for distributed modeling and monitoring: (1) how to evaluate static and dynamic correlations to decompose the whole process into sub-systems under the closed-loop control; (2) how to effectively model static and dynamic characteristics for each sub-system and how to describe the static and dynamic correlations among different sub-systems. To the best of our knowledge, few works in the literature have considered the above problems. To deal with these problems, this paper proposes a sparse slow feature analysis (SSFA) based distributed dynamic monitoring strategy for large-scale closed-loop processes. First, an SSFA algorithm is proposed to concurrently evaluate static and dynamic correlations. Then, based on the SSFA algorithm, an iterative variable subset partition algorithm is designed to decompose the whole process into different subsystems. Finally, SFA and Kernel SFA (KSFA)[24] based distributed modeling and monitoring scheme is developed, where the complete (both static and dynamic) process characteristics within each subsystem along with plant-wide process correlations are modeled and monitored by different methods at different levels. The main contributions of the current work are summarized below.

(1) The effects of control mechanisms are considered for distributed monitoring of the large-scale closed-loop processes. Specifically, an SSFA algorithm is developed and an iterative variable subset partition procedure is designed, which can effectively consider the influences of closed-loop control and divide the complex process into different sub-systems by evaluating static and dynamic correlations.

(2) A two-level modeling and monitoring strategy is proposed, which can effectively describe both the operating conditions and control action from the local and global aspects, thereby obtaining complete process assessment and enhancing monitoring performance.

## 2 SSFA based Distributed Dynamic Modeling and Online Monitoring Method

For closed-loop industrial processes, static distributions reflect steady-state operation patterns while process dynamics reveal control action. Therefore, two critical issues should be considered: First, both the dynamic and static information should be extracted for complete process characteristics evaluation; second, dynamic characteristics should be separated from the static ones because they reveal different aspects of process characteristics. Based on the above consideration, a SSFA based distributed dynamic modeling and monitoring method is proposed for large-scale closed-loop processes in this section, which includes three major steps: 1) A SSFA algorithm is proposed to achieve process decomposition based on the dual evaluation of static and dynamic correlations; 2) SFA and KSFA based distributed modeling strategy is developed to provide explicit explanations of both static and dynamic variations considering local linearity as well as plant-wide nonlinearity; 3) distributed online monitoring method is implemented at two levels to achieve a complete process monitoring. The detailed rationale and implementation procedure are presented as follows.

### 2.1 Sparse slow feature analysis (SSFA) algorithm

As aforementioned, those variables which stay within the same subsystem may have strong static and dynamic correlations thanks to the feedback control mechanism. Therefore, by concurrently evaluating static and dynamic correlations among variables, the whole process can be decomposed into several subsystems. To achieve this purpose, SSFA algorithm is proposed in this paper, which has several advantages. First, it inherits the merit of SFA algorithm that considers both static and dynamic characteristics of data. Second, unlike traditional SFA which involves all the variables, SSFA produces sparse loadings for each slow feature (SF) by setting to zero the coefficients of those unimportant variables. Therefore, by calculating sparse loading vectors, static and dynamic correlations can be concurrently evaluated where important variables are automatically selected. The specific procedure of the SSFA algorithm is then described as follows.

The basic idea of the proposed SSFA is to reformulate SFA as a regression-type optimization problem, where sparse constraint such as lasso or elastic net can be integrated to produce SFs with sparse loadings.

Consider the normalized measurement data $\mathbf{X}(N \times J) = [\mathbf{x}_1, \mathbf{x}_2, ..., \mathbf{x}_J]$, where $\mathbf{x}_j = [x_{j1}, x_{j2}, ..., x_{jN}]^T$ and $J$ denotes the variable dimension. First, to introduce the regression approach to SFA, the penalized minimization problem of traditional SFA algorithm is reformulated as a penalized maximization problem,

$$\begin{aligned} \mathbf{W} &= \arg\max \{\text{tr}(\mathbf{W}^T \mathbf{\Omega} \mathbf{W})\} \\ s.t. \quad &\mathbf{W}^T \dot{\mathbf{\Omega}} \mathbf{W} = \mathbf{I} \end{aligned} \quad (1)$$

where, W denotes the projecting directions which is also called loadings. $\mathbf{\Omega}$ is covariance matrix of process data ($\mathbf{X}$) and $\dot{\mathbf{\Omega}}$ denotes covariance matrix of the first-order derivative of data ($\dot{\mathbf{X}}$). Unlike conventional SFA with the minimization of temporal variations of latent variables (SFs) and constraints on static variations, the reformulated objective function maximizes the static variations of SFs and imposes constraints on temporal SFs. It should be noted that they have different forms but achieve similar results, that is, slowly varying signals can also be calculated by projecting $\mathbf{X}$ on $\mathbf{W}$.

Then, borrowing the similar idea used by Zou et al.[28], the maximization problem in Eq. (1) is transformed into a regression-type problem which aims to calculate **W** best reconstruct the measurements. For each loading **w**, it can be calculated as follows,

$$\mathbf{w} = \arg\min \sum_{i=1}^{N} \left\| \mathbf{x}_i - \mathbf{v}\mathbf{w}^T\mathbf{x}_i \right\|^2 + \lambda \mathbf{w}^T \dot{\mathbf{\Omega}} \mathbf{w} \qquad (2)$$
$$s.t. \quad \mathbf{v}^T \dot{\mathbf{\Omega}} \mathbf{v} = 1$$

where, $N$ denotes the number of samples in **X**, **v** is a dummy variable approximating **w**. Its matrix version can be formulated as follows,

$$\mathbf{W} = \arg\min \left\| \mathbf{X} - \mathbf{X}\mathbf{W}\mathbf{V}^T \right\|^2 + \lambda \sum_{j=1}^{J} \mathbf{w}_j^T \dot{\mathbf{\Omega}} \mathbf{w}_j \qquad (3)$$
$$s.t. \quad \mathbf{V}^T \dot{\mathbf{\Omega}} \mathbf{V} = \mathbf{I}$$

where $\lambda$ is a tuning parameter. However, it is difficult to solve the above optimization problem because of the coupling of variables **W** and **V**, thus Eq.(3) is transformed into the following equivalent problem,

$$\mathbf{W} = \arg\min \left\| \mathbf{X}^*\mathbf{V}^* - \mathbf{X}\mathbf{W} \right\|^2 + \lambda \sum_{j=1}^{J} \mathbf{w}_j^T \dot{\mathbf{\Omega}} \mathbf{w}_j \qquad (4)$$
$$s.t. \quad \mathbf{V}^{*T} \mathbf{V}^* = \mathbf{I}$$

where, $\mathbf{X}^* = \mathbf{X}\mathbf{A}$, $\mathbf{V}^* = \mathbf{A}^T\mathbf{V}$ and $\mathbf{A} = \mathbf{P}^T\mathbf{L}^{1/2}$, $\dot{\mathbf{\Omega}} = \mathbf{A}\mathbf{A}^T$. **L** is the corresponding diagonal matrix of $\dot{\mathbf{\Omega}}$ and **P** satisfies that $\mathbf{P}^T\mathbf{P} = \mathbf{I}$. For ease of reading, the proof is presented in Appendix A to show the equivalence between Eq.(3) and Eq.(4).

**Remark** It is worth mentioning that the norm used for vector is $l2$-norm and for matrix is $F$-norm. In the following part of this work, ‖ ‖ is used to represent both of them without declaration.

So far, we have established the connection between SFA and regression. Next, in order to obtain sparse loadings which are in fact the regression coefficient vectors, an $l1$-norm penalty is applied to Eqs.(3) and (4). Then, our sparse slow feature analysis (SSFA) criterion is defined as below,

$$\mathbf{w}_j = \arg\min \|\mathbf{X}^*\mathbf{v}_j^* - \mathbf{X}\mathbf{w}_j\| + \lambda \mathbf{w}_j^T \dot{\boldsymbol{\Omega}} \mathbf{w}_j + \lambda_{1,j} \|\mathbf{w}_j\|_1$$
$$s.t. \quad \mathbf{v}_j^{*T}\mathbf{v}_j^* = 1 \tag{5}$$

Its matrix version can be formulated as

$$\mathbf{W} = \arg\min \|\mathbf{X}^*\mathbf{V}^* - \mathbf{X}\mathbf{W}\|^2 + \lambda \sum_{j=1}^{J} \mathbf{w}_j^T \dot{\boldsymbol{\Omega}} \mathbf{w}_j + \sum_{j=1}^{J} \lambda_{1,j} \|\mathbf{w}_j\|_1$$
$$s.t. \quad \mathbf{V}^{*T}\mathbf{V}^* = \mathbf{I} \tag{6}$$

where $\dot{\boldsymbol{\Omega}}$ denotes the temporal covariance matrix which is positive definite, and $\lambda$ and $\lambda_{1,j}$ are nonnegative tuning parameters. The $l1$-norm penalty results in sparsity when $\lambda_{1,j}$ is large enough.

In order to solve the above optimization problem, an iterative algorithm is adopted here. If $\mathbf{V}^*$ is fixed, the above problem is in fact a generalized elastic net problem[28], which can be transformed to a lasso-type problem (see the proof in Appendix B) and readily solved by LARS-EN algorithm[29]. If $\mathbf{W}$ is fixed, then the above optimization problem is boiled down to the following,

$$\mathbf{V}^* = \arg\min \|\mathbf{X}^*\mathbf{V}^* - \mathbf{X}\mathbf{W}\|^2 = \arg\min \|\mathbf{X}^* - \mathbf{X}\mathbf{W}\mathbf{V}^{*T}\|^2$$
$$s.t. \quad \mathbf{V}^{*T}\mathbf{V}^* = \mathbf{I} \tag{7}$$

Then, Eq.(7) can be easily solved according to Reduced Rank Procrustes Rotation Theorem[28].

Therefore, the SSFA algorithm can be presented as follows.

**Step 1**. Let $\mathbf{V}^*$ start from $\mathbf{W}^*$ which is the loading matrix of conventional SFA on $\mathbf{X}^*$.

**Step 2**. Given a fixed $\mathbf{V}^* = [\mathbf{v}_1, \mathbf{v}_2, ..., \mathbf{v}_J]$, solve the following generalized elastic net problem for $j = 1, 2, ..., J$

$$\mathbf{w}_j = \arg\min \|\mathbf{X}\mathbf{v}_j^* - \mathbf{X}\mathbf{w}_j\|^2 + \lambda \mathbf{w}_j^T \dot{\boldsymbol{\Omega}} \mathbf{w}_j + \lambda_{1,j} \|\mathbf{w}_j\|_1$$

**Step 3**. Given a fixed **W**, perform SVD of $\mathbf{X}^{*}\mathbf{X}\mathbf{W} = \mathbf{Q}\mathbf{D}\mathbf{R}^{\mathrm{T}}$ according to Reduced Rank Procrustes Rotation Theorem, and then update $\mathbf{V}^{*}$ by $\mathbf{V}^{*} = \mathbf{Q}\mathbf{R}^{\mathrm{T}}$.

**Step 4**. Repeat Steps 2 to 3 until $\mathbf{V}^{*}$ and **W** convergence or the maximum iteration number is reached.

**Step 5.** Normalization: $\mathbf{w}_{j} = \dfrac{\mathbf{w}_{j}}{\left\|\mathbf{w}_{j}^{\mathrm{T}}\dot{\mathbf{\Omega}}\mathbf{w}_{j}\right\|}$.

Note that SSFA algorithm derives from the traditional SFA algorithm, the loadings (**W***) of SFA are selected as the initial point of **V*** to reduce the iteration steps. In summary, the proposed SSFA algorithm extracts sparse loadings on which the output signal (SF) varies as slowly as possible. Here, **W** is the sparse loading matrix in which each column denotes a sparse loading. For each sparse loading, the coefficients of irrelevant variables are reduced to zero values, denoting that each SF obtained by SSFA is in fact a linear combination of some highly correlated variables rather than the whole input variables. Those variables corresponding to non-zero linear combination coefficients are associated to each SF. Moreover, considering that SSFA simultaneously explores both static covariance matrix and temporal covariance matrix of process data, variables associated to each SF have both strong static and dynamic correlations. Besides, the value of temporal variation (slowness) for each SF reflects the correlation extent of associated variables. Since the calculated SFs are automatically arranged in a descending order according to their temporal variations, those variables associated to the first SF have the strongest static and dynamic correlations. The property of SSFA algorithm will be illustrated in Case Study Section shortly.

**2.2 SSFA based iterative variable subset partition algorithm for process decomposition**

As SSFA can select those variables with strong static and dynamic correlations, a SSFA based iterative variable subset partition algorithm is proposed for process decomposition in this section. In each iterative procedure, a group of variables can be determined according to the first sparse loading. Besides, considering that different subsets should be as diverse as possible to reduce the redundancy among different control loops as well as to capture complementary process information, the earlier determined variable subset is excluded from original data before the partition of the next subset. The specific procedure is presented below.

**Step 1.** Perform SSFA on the input data $\mathbf{X}$ to obtain the first sparse loading $\mathbf{w}$. Archive those variables $\mathbf{X}_s(N \times J_s)$ (the subscript $s$ denotes those selected variables) indicated by non-zero coefficients in $\mathbf{w}$ into the first subset, called S&DL subset (static and dynamic linear subset). Then, perform SFA on them to obtain the slow features $\mathbf{S}_s(N \times J_s)$ with all features retained and calculate the slowness vector ($\mathbf{SL}_s$) for $\mathbf{S}_s$, which is indicated as

$$\mathbf{SL}_s = \left(\mathbf{SL}_{s,1}, \mathbf{SL}_{s,2}, \ldots, \mathbf{SL}_{s,N}\right)^T \text{ and } \mathbf{SL}_{s,n} = \frac{\dot{\mathbf{s}}_{s,n}^T \dot{\mathbf{\Xi}}_s^{-1} \dot{\mathbf{s}}_{s,n}}{\mathbf{s}_{s,n}^T \mathbf{\Xi}_s^{-1} \mathbf{s}_{s,n}}$$

, where $\dot{\mathbf{s}}_{s,n}$ and $\mathbf{s}_{s,n}$ denotes the $n$th row of $\dot{\mathbf{S}}_s$ and $\mathbf{S}_s$ respectively, $\dot{\mathbf{\Xi}}_s$ and $\mathbf{\Xi}_s$ are the covariance matrixes of $\dot{\mathbf{S}}_s$ and $\mathbf{S}_s$.

**Step 2.** Remove $\mathbf{X}_s$ from the current input data $\mathbf{X}$ to generate the new input data $\mathbf{X}_r$. Perform SSFA on $\mathbf{X}_r$ to select those new statically and dynamically linear correlated variables ($\mathbf{X}_s^*$). Subsequently, calculate new slow features $\mathbf{S}_s^*$ and new slowness vector $\mathbf{SL}_s^*$ using the same way as described in Step 1.

**Step 3.** Conduct paired-nonparametric test for $\mathbf{SL}_s^*$ and $\mathbf{SL}_s$ to check whether there is significant difference between them. If $\mathbf{SL}_s^*$ and $\mathbf{SL}_s$ show no significant difference, it means $\mathbf{S}_s^*$ and $\mathbf{S}_s$ have comparable slowness. It means that those variables in $\mathbf{X}_s^*$ have strong static and dynamic correlations and thus $\mathbf{X}_s^*$ can be regarded as the S&DL subset. Then remove $\mathbf{X}_s^*$ from the current input data and go to Step 4. Otherwise, it means that $\mathbf{S}_s^*$ does not vary so slowly as $\mathbf{S}_s$, that is, the related variables are statically and dynamically linear uncorrelated. Therefore, $\mathbf{X}_s^*$ should be excluded from the S&DL subsets and the procedure should be stopped.

**Step 4**. Repeat Step 2 and Step 3 until no more S&DL subsets can be determined. Then, the remaining variables are statically and dynamically linear uncorrelated, forming the S&DNL subset (static and dynamic nonlinear subset).

The flowchart of variable subset partition procedure is presented in Figure 1, where, the input is measured process data and the output are those separated subsets comprising certain S&DL subsets and one S&DNL subset. For each S&DL subset, SFs can be calculated through linear mapping functions. For the S&DNL subset, SFs cannot be extracted by linear mapping functions because variables are linear uncorrelated from either static or dynamic aspect.

### *Discussion*

(1). Quadratic penalized parameter $\lambda$ and $l$1-norm penalized parameter $\lambda_{1,j}$

The quadratic penalized parameter $\lambda$ is used to smooth the variable coefficients among correlated variables. As a result, the quadratic penalty can be regarded as the consideration of variable correlations when $\lambda > 0$. Experience shows that the output of SSFA algorithm does not change much as $\lambda$ varies, and thus $\lambda$ is usually chosen to be a small positive number. The $l1$-norm penalized parameter $\lambda_{1,j}$ plays an important role in SSFA algorithm, which determines the sparsity (the number of non-zero coefficients) of SSFA algorithm and the SSFA based variable subset partition. Generally, a larger $\lambda_{1,j}$ would result in higher sparsity and a smaller $\lambda_{1,j}$ achieves the opposite. Since the LARS-EN algorithm we utilized can effectively produce a whole sequence of sparse loadings and the corresponding values of $\lambda_{1,j}$ for each SF, $\lambda_{1,j}$ is usually determined to be the one with the best fitting error.

(2). Slowness vector ($\mathbf{SL}$)

As mentioned above, the slowness vector for selected variables $\mathbf{X}_s$ is defined as follows:

$$\mathbf{SL}_s = \left(\mathbf{SL}_{s,1}, \mathbf{SL}_{s,2}, ..., \mathbf{SL}_{s,N}\right)^{\mathrm{T}} \text{ and } \mathbf{SL}_{s,n} = \frac{\dot{\mathbf{s}}_{s,n}^{\mathrm{T}} \dot{\mathbf{\Xi}}_s^{-1} \dot{\mathbf{s}}_{s,n}}{\mathbf{s}_{s,n}^{\mathrm{T}} \mathbf{\Xi}_s^{-1} \mathbf{s}_{s,n}}$$

where $N$ is the number of samples and $n$ denotes the $n$th sample, $\dot{\mathbf{s}}_{s,n}$ and $\mathbf{s}_{s,n}$ denote the $n$th row of $\dot{\mathbf{S}}_s$ and $\mathbf{S}_s$, $\dot{\mathbf{\Xi}}_s$ is the covariance matrix of $\dot{\mathbf{S}}_s$ and $\mathbf{\Xi}_s$ is the covariance matrix of $\mathbf{S}_s$ which is in fact an identity matrix. It is clear that $\mathbf{SL}_{s,n}$ captures the slowness extent at each sampling time and thus $\mathbf{SL}_s$ reveals the overall slowness. Therefore, by checking the differences of $\mathbf{SL}$ among different subsets, the extents of slowness of the different subsets can be compared. During the above variable subset partition procedure, when a subset is initiated by the concerned sparse loading, $\mathbf{SL}$ is calculated and compared with that of the former selected S&DL subset. If the current $\mathbf{SL}$ shows significant difference from the former one, it denotes that the slowness extent of the current subset has been reduced significantly and thus the current subset should be excluded from S&DL subsets. Here, considering $\mathbf{SL}$ does not follow Gaussian distribution, the paired-nonparametric test approach[30] is utilized to test the difference between different **SLs**.

### 2.3 Two-level based distributed monitoring system

Until now, the measured variables have been separated into several S&DL subsets and one S&DNL subset. To capture both the linear characteristics and nonlinear characteristics from both the static and dynamic aspects, a two-level based distributed modeling strategy is proposed in this section, where in the first modeling level, SFA is performed on each S&DL subset and in the second level, Kernel SFA is adopted upon those super samples (variables in S&DNL subset and slow features extracted from each S&DL subset). The specific modeling procedure is presented as follows.

For the first level modeling, SFA is performed on each S&DL subset $\mathbf{X}_k (k=1,2,...,K)$

$$\begin{aligned} \mathbf{W}_k &= \arg\min\{\operatorname{tr}(\mathbf{W}^T \dot{\mathbf{\Omega}}_k \mathbf{W})\} \\ s.t. \quad & \mathbf{W}^T \mathbf{W} = \mathbf{I} \\ \mathbf{S}_k &= \mathbf{X}_k \mathbf{W}_k \\ \dot{\mathbf{S}}_k &= \dot{\mathbf{X}}_k \mathbf{W}_k \end{aligned} \quad (8)$$

where, the subscript $k$ denotes the $k$th slow subset, $\mathbf{W}_k$ contains all the loadings, $\mathbf{S}_k$ and $\dot{\mathbf{S}}_k$ include all the SFs and temporal SFs without dimension reduction. For dimension reduction and de-noising, it is desired that system information is preserved whilst noises are removed. In practice, the differences between system characteristics and noises are mainly manifested in two aspects: first, system variation may vary more slowly than noises; besides, system variations may be larger than that of noises. Based on such consideration, the preserved slow features should have large variations and small temporal variations. Then, the criterion of selecting $M$ system slow components is defined as follows. That is, the ratio of dynamic variations to static variations of retained SFs should be smaller than the largest value from the measurement variables. This criterion is described as

$$M = \text{card}\left\{\dot{\mathbf{s}}_j \mid \frac{\dot{\mathbf{s}}_j^T \dot{\mathbf{s}}_j}{\mathbf{s}_j^T \mathbf{s}_j} \leq \max\left(\frac{\dot{\mathbf{x}}_j^T \dot{\mathbf{x}}_j}{\mathbf{x}_j^T \mathbf{x}_j}\right)\right\} \quad (9)$$

where, $\dot{\mathbf{s}}_j$ denotes the first order derivative of $\mathbf{s}_j$ and it can be calculated as $\dot{\mathbf{s}}_j(n) = \mathbf{s}_j(n) - \mathbf{s}_j(n-1)$, where $n$ denotes the $n$th sample; and card{} denotes the number of elements in a certain set. As a result, in each S&DL subset, slow features can be divided into two parts, system slow features $\mathbf{S}_{k,s}$ which carry slowly-varying system information, and residual slow features $\mathbf{S}_{k,f}$ containing fast-varying noises. Subsequently, the related monitoring statistics can be calculated,

$$\begin{aligned} T_{k,\bullet}^2 &= \mathbf{s}_{k,\bullet}^T \mathbf{s}_{k,\bullet} \\ D_{k,\bullet}^2 &= \dot{\mathbf{s}}_{k,\bullet}^T \dot{\boldsymbol{\Omega}}_k^{-1} \dot{\mathbf{s}}_{k,\bullet} \end{aligned} \quad (10)$$

where, the subscript • is used to indicate both system slow features and residual slow features, $\mathbf{s}_{k,\bullet}$ and $\dot{\mathbf{s}}_{k,\bullet}$ are row vectors from the residual slow feature $\mathbf{S}_{k,\bullet}$ and temporal residual slow feature $\dot{\mathbf{S}}_{k,\bullet}$, $\dot{\boldsymbol{\Omega}}_{k,\bullet}$ denotes the covariance matrix of $\dot{\mathbf{S}}_{k,\bullet}$. The static monitoring indexes $T_{k,s}^2$ and $T_{k,f}^2$ capture linear operating conditions in each S&DL subset whilst the dynamic monitoring indexes $D_{k,s}^2$ and $D_{k,f}^2$ reveal the local control action. The control limits of the above statistics can be calculated by kernel density estimation (KDE) method[31], in which the control limits with significant level $\alpha\%$ is determined as the point whose cumulative probability is close to $\alpha\%$.

For the second-level modeling, super samples ($\mathbf{X}_{sp}$) are firstly constructed by combining system slow features $\mathbf{S}_{k,s}$ of each S&DL subset with variables in S&DNL subset. The reason to collect system slow features rather than residual slow features is that: system slow features contain major information and residual slow features may be regarded as noises. Therefore, to avoid loss of system information, system slow features are collected to form super samples. Then, KSFA is adopted to model the plant-wide static and dynamic process distributions. The detailed modeling procedure is presented as below.

Given the super sample matrix $\mathbf{X}_{sp} = [\mathbf{x}_{sp,1}, \mathbf{x}_{sp,2}, ..., \mathbf{x}_{sp,J}]$, for each super sample $\mathbf{x}_{sp,j}(j=1,2,...,J)$, we can obtain the super SFs $\mathbf{s}_{sp}^{\mathrm{T}} = [s_{sp,1}, s_{sp,2}, ..., s_{sp,J}]$. To select $M$ system slow features, the criterion described in Eq.(9) can also be adopted and modified here,

$$M = \mathrm{card}\left\{\dot{\mathbf{s}}_j \mid \frac{\dot{\mathbf{s}}_{sp,j}^{\mathrm{T}}\dot{\mathbf{s}}_{sp,j}}{\mathbf{s}_{sp,j}^{\mathrm{T}}\mathbf{s}_{sp,j}} \leq \max\left(\frac{\Phi(\dot{\mathbf{x}}_j)^{\mathrm{T}}\Phi(\dot{\mathbf{x}}_j)}{\Phi(\mathbf{x}_j)^{\mathrm{T}}\Phi(\mathbf{x}_j)}\right)\right\}$$
$$= \mathrm{card}\left\{\dot{\mathbf{s}}_j \mid \frac{\dot{\mathbf{s}}_{sp,j}^{\mathrm{T}}\dot{\mathbf{s}}_{sp,j}}{\mathbf{s}_{sp,j}^{\mathrm{T}}\mathbf{s}_{sp,j}} \leq \max\left(\frac{\dot{\tilde{\mathbf{K}}}_{sp,jj}}{\tilde{\mathbf{K}}_{sp,jj}}\right)\right\}$$
(11)

where, the subscript $j$ denotes the $j$th element, $\dot{\tilde{\mathbf{K}}}_{sp,jj}$ denotes the $j$th diagonal element of super temporal kernel matrix $\dot{\tilde{\mathbf{K}}}_{sp}$, and $\tilde{\mathbf{K}}_{sp,jj}$ represents the $j$th diagonal element of super kernel matrix $\tilde{\mathbf{K}}_{sp}$. As a result, the super SFs can also be separated into super system slow features ($\mathbf{S}_{sp,s}$) and super residual slow features ($\mathbf{S}_{sp,f}$). For process monitoring, two groups of statistics are calculated as below,

$$T_{sp,\bullet}^2 = \mathbf{s}_{sp,\bullet}^{\mathrm{T}}\mathbf{s}_{sp,\bullet}$$
$$D_{sp,\bullet}^2 = \dot{\mathbf{s}}_{sp,\bullet}^{\mathrm{T}}\dot{\mathbf{\Omega}}_{sp}^{-1}\dot{\mathbf{s}}_{sp,\bullet}$$
(12)

where, the static indexes $T_{sp,\bullet}^2$ capture global nonlinear operating conditions and the dynamic indexes $D_{sp,\bullet}^2$ summarize the global process dynamics revealing the overall control action.

The proposed two-level distributed modeling method is illustrated in Figure 2. In this way,

steady-state operation conditions and the effects of the control mechanism are separately described at different levels respectively, where the first-level monitoring system focuses on local linear patterns and the second-level model mainly describes plant-wide nonlinear behaviors.

### 2.4 Two-level based distributed online monitoring strategy

Whenever new observations are available, each of the samples is first processed by the normalization information of training data and then arranged according to the variable subset partition result. Subsequently, operation conditions and dynamic behaviors are checked based on the established distributed monitoring system. The detailed procedure is implemented as follows.

Given the new sample $\mathbf{x}_{new}$, distributed SFA models $\mathbf{W}_k$ are utilized to detect the static and dynamic linear patterns in each S&DL subset,

$$\begin{aligned}\mathbf{s}_{new,k,\bullet} &= \mathbf{W}_{k,\bullet}^{\mathrm{T}}\mathbf{x}_{new,k} \\ \dot{\mathbf{s}}_{new,k,\bullet} &= \mathbf{W}_{k,\bullet}^{\mathrm{T}}\dot{\mathbf{x}}_{new,k}\end{aligned} \quad (13)$$

where, the subscript $\bullet$ is used to indicate both system slow features and residual ones, and $\mathbf{s}_{new,k,\bullet}$ and $\dot{\mathbf{s}}_{k,new,\bullet}$ are system feature and temporal system feature associated to $\mathbf{x}_{new,k}$. Two groups of monitoring statistics are then calculated,

$$\begin{aligned}T_{new,k,\bullet}^2 &= \mathbf{s}_{new,k,\bullet}^{\mathrm{T}}\mathbf{s}_{new,k,\bullet} \\ D_{new,k,\bullet}^2 &= \dot{\mathbf{s}}_{new,k,\bullet}^{\mathrm{T}}\dot{\boldsymbol{\Omega}}_{new,k}^{-1}\dot{\mathbf{s}}_{new,k,\bullet}\end{aligned} \quad (14)$$

where, the static index captures steady-state operating patterns and the dynamic index reveals control action of the closed-loops. It is noted that if any subset issues alarming static indexes, the local operating pattern has changed, which might be caused by process faults or normal operating state changes. Then, dynamic indexes are checked. If corresponding dynamic indexes stay beyond the control limits, process fault occurs; otherwise, process might shift from the reference operating condition to a new one and is not in a faulty condition.

To further check the global process characteristics, the second-level monitoring procedure is conducted. First, system slow features ($\mathbf{s}_{new,k,s}$) in each S&DL subset are collected with those variables in the S&DNL subset to form the super sample ($\mathbf{x}_{new,sp}$). Then, KSFA is performed on $\mathbf{x}_{new,sp}$ to concurrently monitor the static and dynamic nonlinear patterns,

$$\begin{aligned} T^2_{new,sp,\bullet} &= \mathbf{s}^{\mathrm{T}}_{new,sp,\bullet}\mathbf{s}_{new,sp,\bullet} \\ D^2_{new,sp,\bullet} &= \dot{\mathbf{s}}^{\mathrm{T}}_{new,sp,\bullet}\Omega^{-1}_{sp,\bullet}\dot{\mathbf{s}}_{new,sp,\bullet} \end{aligned} \qquad (15)$$

where, $\mathbf{s}_{new,sp,\bullet}$ is the new super components obtained by the kernel projection of the super sample $\mathbf{x}_{new,sp}$, and $\dot{\mathbf{s}}_{new,sp,\bullet}$ denotes the new temporal super components. If both the static monitoring indexes and dynamic indexes alarm, the process fault with nonlinear characteristics is detected. If only static statistics exceed the control limits, conclusion could be made that process experiences normal operating condition change and therefore not in a faulty condition.

During the above monitoring procedure, static and dynamic process characteristics are checked in both subset-wise and plant-wide. The interpretations of monitoring indexes are summarized in Table I. Moreover, the monitoring policy can be summarized as below.

*<u>Monitoring Policy</u>*

*For each subset (the first level)*

(1). If both $T^2_{new,k,\bullet}$ and $D^2_{new,k,\bullet}$ exceed the control limits, it means that the process deviates from its predefined operating conditions and the local disturbance cannot be compensated by control action. Therefore, process is out of control and a fault happens in the concerned subset.

(2). If $D^2_{new,k,\bullet}$ first alarms and then goes back to normal after $T^2_{new,k,\bullet}$ alarms, it means that the local disturbance can be compensated by control action and thus the local process is still under control at a new steady operating condition which has not been included in the modeling data.

(3). If neither $T^2_{new,k,\bullet}$ nor $D^2_{new,k,\bullet}$ is out of control, it means that the process stays at a predefined normal condition in the *k*th subset.

Based on the monitoring results of first level, different indexes of the second-level monitoring system are used to further check the process status:

*For global process (the second level)*

(1). When the local process has not been affected,

 a) If both $T^2_{new,sp,\bullet}$ and $D^2_{new,sp,\bullet}$ are out of control, it means that the whole process deviates from its reference operating conditions and the disturbance cannot be compensated, indicating that the control loops cannot well coordinate the relationships among different subsets and thus a process fault happens.

 b) If $D^2_{new,sp,\bullet}$ first alarms and then returns to normal after $T^2_{new,sp,\bullet}$ alarms, it means that a new global static nonlinear relationship has been identified where the disturbance can be well compensated by control action, indicating a new plant-wide operating condition that has not been included in the historical data.

 c) If neither $T^2_{new,sp,\bullet}$ nor $D^2_{new,sp,\bullet}$ is out of control, the plant-wide process remain in

the predefined normal operating conditions.

(2). When the local process has been affected,

a) If both $T^2_{new,sp,\bullet}$ and $D^2_{new,sp,\bullet}$ are out of control, it denotes that relationships among different subsets cannot be well coordinated due to either the deviations of local process operating conditions or local process faults, indicating that the fault makes whole process out of control.

b) If $D^2_{new,sp,\bullet}$ goes back to normal after $T^2_{new,sp,\bullet}$ alarms, it indicates that the local disturbance(operating condition change or process fault) can be compensated by control action and thus the whole process enters into a new operating conditions that has not been included in the historical data.

c) If neither $T^2_{new,sp,\bullet}$ nor $D^2_{new,sp,\bullet}$ is out of control, it means that this disturbance (operating condition change or process fault) just affects local process behaviors.

## 3 Case Study

### 3.1 Tennessee Eastman (TE) benchmark process

The proposed method is applied to the well-known TE benchmark process[32], which consists of five major units: reactor, condenser, compressor, separator and stripper. A PI-control strategy is implemented to control the process. Therefore, variables collected from TE process can be divided into two blocks: the XMV block with 12 manipulated variables (XMV #1~#12) and XMEAS with 41 measured variables (XMEAS #1~#41). Though the process scale of TE benchmark may not be large, it can be used to illustrate the effectiveness of the proposed SSFA algorithm and SSFA based variable subset partition algorithm. In this study, only part of manipulated variables (XMV #1~#11) and measured variables (XMEAS #1~#22) sampled every 3 minutes are chosen as the input. One normal data set is generated as training data, which comprises 500 samples. Testing data includes normal data and two types of faults and each of them contains 480 samples.

Fault #1: IDV(10), random variation in C feed temperature. In this case, the random variation in C feed temperature results in a persistent fluctuation of stripper temperature from the beginning to the end of process. Correspondingly, the stripper steam valve is manipulated to alleviate the variations of stripper temperature, thus the operation conditions and process dynamics are both affected.

Fault #2: IDV(5), step variation in condenser cooling water inlet temperature from the beginning of the process. In this case, the step change of condenser cooling water inlet temperature mainly results in a step increase of the condenser cooling water flow rate, which subsequently causes an increase in temperature of the separator, and finally increases the separator cooling water outlet temperature. However, due to the control loops, this disturbance can well be compensated and the temperature in separator returns to normal in about 10 hours (200 samples).

First, a slowness index $sl$ is calculated for each SF by SSFA and SFA, which is indicated as $sl = \frac{\dot{\mathbf{s}}^T \dot{\mathbf{s}}}{\mathbf{s}^T \mathbf{s}} = \dot{\mathbf{s}}^T \dot{\mathbf{s}}$, where $\dot{\mathbf{s}}$ denotes the temporal SF, and $\mathbf{s}$ denotes SF which is normalized to have zero mean and unit variation. The calculation results are presented in Figure 3, where the y-axis denotes the slowness index and x-axis denotes the order of projection directions. Besides, $\lambda = 1.5$ and each $\lambda_{1,j}$ is determined as the one with smallest slowness which is not presented here. From Figure 3 it can be seen that SFs of SSFA have the similar slowness trend (monotonously increase) as that of traditional SFA algorithm. Moreover, SSFA can obtain slower SFs because statically and dynamically correlated variables are picked up for extracting SFs, indicating a more reliable variable selection mechanism. Table II shows the properties of the first five sparse loadings obtained by SSFA. It is noted that only a small numbers of variables are selected for each SF, denoting high sparsity. The coefficients of unimportant variables are decreased to zero for each SF, which makes interpretation clearer and easier.

Subsequently, SSFA based variable subset partition procedure is implemented, in which the typical paired-nonparametric test, sign-rank test[27] is used to evaluate the differences of slowness extents (indicated by **SL**) of different subsets. Five S&DL subsets and one S&DNL subset are separated, in which the partition results of five S&DL subsets are presented in Table III. To further demonstrate the effectiveness of the proposed variable subset partition algorithm, some intermediate results are presented in Figure 4. Figure 4(a) shows the probability distributions of **SL** in six determined S&DL subsets (Subsets #1~#6) in which one S&DL subset (Subset #6) should be excluded since its distribution has changed significantly in comparison with those of Subsets #1~#5. Figure 4(b) presents the quantitative difference test result, where the y-axis denotes the significance level and x-axis indicates the sequence of partitioned subsets. It can be seen that the significance level shows a decrease trend when more subsets are determined. For Subset #6, the significant level stays below 5% (red dashed line), which means that the hypothesis that Subset #6 has similar slowness extent as the former one should be rejected. As a result, Subset #6 cannot be regarded as the S&DL subset and the partition procedure should be terminated.

In order to further show the effectiveness of the proposed SSFA based variable subset partition method, the online monitoring results of the proposed method are compared with traditional methods (such as PCA, SFA[25] and H-PCA-KPCA[19]). First, for those normal testing data, all methods show the comparable performance to accommodate normal variations, where the values of FAR are less than 6% for all concerned monitoring indexes. For abnormal testing data, monitoring results are presented in Figure 5 and Figure 6.

For Fault #1, Figure 5 (a) presents the monitoring results of SFA. It is noted that both the static and dynamic indexes issue persistent alarms, denoting that a real process fault occurs and process is problematic. However, a large number of samples stay below the control limits during the time period (200, 400), resulting in confusing conclusion that process is still under control and no maintenance is required. Figure 5 (b) shows that PCA is insensitive to this anomaly since numerous alarms are missed. As shown in Figure 5 (c), H-PCA-KPCA can correctly identify the abnormal static variations. However, no information could be revealed about process dynamics and thus it cannot judge whether process swifts to a new operating conditions or a fault happens. Figure 5 (d) shows the monitoring results of the proposed method at different levels. It is clear that both the static indexes and dynamic indexes at the second level go beyond the control limits, thereby delivering reliable information that process dynamics are disrupted and maintenance is required. Besides, detailed fault information can be speculated from the first level. That is, Fault #1 mainly affects variables of Subset #1

(stripper temperature, stripper steam flow and the corresponding control variable) and then propagates to other variable subsets with delayed alarms (shown in Subsets #2 and #3), leading to abnormal static and dynamic behaviors.

For Fault #2, as shown in Figure 6 (a), the static indexes of SFA algorithm issue persistent alarms from the beginning to the end whilst the dynamic index first alarms and then returns to normal, denoting that process goes through a normal operating condition change. However, consistently missed alarms are found in $S^2$, which would also deliver confusing information. Figure 6 (b) shows that the operating condition deviation cannot be detected by PCA after the 200th sample. The monitoring result of H-PCA-KPCA is presented in Figure 6 (c), where abnormal static variations can be well detected but process dynamics is missed. Figure 6 (d) shows the monitoring results of the proposed distributed SSFA method. It is could be derived that process deviates from its predefined operating condition at the beginning and then enters a new steady operating condition after the 200th sample. Moreover, it should be mentioned that variables within all the S&DL subsets have transients that settle in about 10 hours (200samples) and finally go back around their set points, which denotes that the deviation of process operating condition may be caused by those variables beyond the S&DL subsets. This speculation agrees with the fact that the significant effect of the fault is to induce a step change in cooling water flow rate (Variable #32), which is included in the S&DNL subset.

**3.2 1000 MW ultra-supercritical thermal power unit**

The 1000 MW ultra-supercritical thermal power unit is an advanced power generation technique with high plant efficiency, high coal utilization and low emission. In general, the

1000 MW ultra-supercritical thermal power unit is a large-scale, highly complex generation system with up to 30 MPa steam pressure and 600 °C steam temperature[33]. Therefore, the 1000 MW ultra-supercritical thermal power unit is utilized here to illustrate effectiveness of the proposed distributed approach for monitoring practical large-scale closed-loop industrial process. Here, the data are collected from Jiahua thermal power plant, where 159 variables are selected and the normal training data are collected under certain steady operating conditions. Besides, two testing data sets are generated, in which one is collected during the period that process shifts from the reference state to a new operating state and the other one is a real process fault. Specific information of testing data is presented as follows:

Case #1: Nominal operating condition change. In this case, process deviates from its reference operating condition around the 100$^{th}$ sample and enters a new steady state around the 400$^{th}$ sample.

Case #2: Real fault. It induces an increased pressure drop of inner cooling water of condenser water chamber from the beginning of process.

First, based on the normal training data, five S&DL subsets are determined by the proposed SSFA based variable subset partition algorithm. The specific partition result can be seen in Table IV, where 7, 9, 8, 19 and 10 variables are included in these five subsets, respectively. Meanwhile, the coefficients of the first sparse loadings of SSFA for different S&DL subsets are presented in Figure 7, which tells informative physical interpretations for SFs. For example, as shown in the top sub-plot of Figure 7, those variables with significantly large coefficients are Variable #40 (temperature of radial bearing of front pump A), Variable

#49 (mechanical seal water temperature of front pump A) and Variable #59 (motor coil temperature of front pump A), which are temperatures of different parts of pump A. Generally, the increase of motor coil's temperature would lead to an increment of the temperature of radial bearing. Then, in order to prevent overheating, the mechanical seal water will absorb the heat of radial bearing. Therefore, they may have strong static and dynamic correlations and their combination may generate SFs.

Then, the monitoring results for Case #1 and Case #2 are reported. First, the first system slow features of each S&DL subset for Case #1 and Case #2 are presented in Figure 8. For Case #1, Figure 8(a) demonstrates that these system slow features reveal the system trend that shifts from the reference operating condition to a new one, which agrees well with the real fact. For Case #2, Figure 8(b) shows that some system slow features have become varying fast and thus process may oscillate and go beyond control. Figure 9 presents the monitoring charts of the proposed distributed SSFA algorithm for Case #1 and Case #2. Figure 9(a) presents the monitoring results for Case #1, from which it is clear that the static indexes issue persistent alarms after the $100^{th}$ sample, denoting a deviation from the reference operating condition; meanwhile, the alarms of dynamic indexes only last for a period of time (about 300 samples) and then are eliminated, which indicates that process is finally under control. Therefore, it could be concluded that process experiences a normal operating condition change, which agrees with the real fact. As presented in Figure 9(b), both the static and dynamic indexes issue persistent alarms for Case #2. Therefore, conclusion could be drawn that a real fault happens and process is problematic.

## 4 Conclusion

In this paper, a distributed dynamic method is proposed for modeling and monitoring large-scale closed-loop industrial processes. First, SSFA algorithm is proposed and it can effectively select those statically and dynamically correlated variables. Based on SSFA, the whole process has been decomposed into several subsets by the indication of sparse loadings. Subsequently, distributed SFA models are developed at two levels. At the first level, SFA explores static and dynamic linear patterns within each S&DL subset; at the second level, KSFA serves as the global model and captures plant-wide nonlinearity from both the static and dynamic aspects. For online application, four groups of indices have been proposed at two levels, which check the different aspects of process characteristics. Unlike traditional distributed approaches, the proposed method considers the influence of closed-loop actions revealed by process dynamics. As a result, it can well distinguish real faults from normal operating condition change, which is of great importance for practical applications. The proposed method is verified by TE benchmark data and a real industrial application, the 1000 MW ultra-supercritical thermal power unit.

**Acknowledgement**

This work is supported by the NSFC-Zhejiang Joint Fund for the Integration of Industrialization and Informatization (No.U1709211) and the National Natural Science Foundation of China (No. 61433005), and Natural Science and Engineering Research Council of Canada, and the Research Project of the State Key Laboratory of Industrial Control Technology, Zhejiang University, China (ICT1802), and 111 project under Grant B17040.

**Appendix A**

*Proof*

Since $\dot{\mathbf{\Omega}}$ is a symmetric matrix, it satisfies that

$$\dot{\mathbf{\Omega}} = \mathbf{P}\mathbf{L}\mathbf{P}^{\mathrm{T}} = \mathbf{P}\mathbf{L}^{1/2}\mathbf{L}^{1/2}\mathbf{P}^{\mathrm{T}} = \mathbf{A}\mathbf{A}^{\mathrm{T}} \tag{A.1}$$

where, $\mathbf{L}$ is the corresponding diagonal matrix of $\dot{\mathbf{\Omega}}$ and $\mathbf{P}$ is a unit orthogonal matrix satisfying $\mathbf{P}^{\mathrm{T}}\mathbf{P} = \mathbf{I}$.

Then, we have

$$\begin{aligned}
\min \|\mathbf{X} - \mathbf{X}\mathbf{W}\mathbf{V}^{\mathrm{T}}\| &\Leftrightarrow \min \|\mathbf{X} - \mathbf{X}\mathbf{W}\mathbf{V}^{\mathrm{T}}\| \|\mathbf{P}\mathbf{L}^{1/2}\| \|\mathbf{L}^{1/2}\mathbf{P}^{\mathrm{T}}\mathbf{V}\| \\
&= \min \|(\mathbf{X} - \mathbf{X}\mathbf{W}\mathbf{V}^{\mathrm{T}})\mathbf{P}\mathbf{L}^{1/2}\mathbf{L}^{1/2}\mathbf{P}^{\mathrm{T}}\mathbf{V}\| \\
&= \min \|(\mathbf{X}\mathbf{P}\mathbf{L}^{1/2})(\mathbf{L}^{1/2}\mathbf{P}^{\mathrm{T}}\mathbf{V}) - \mathbf{X}\mathbf{W}\| \\
&= \min \|\mathbf{X}^{*}\mathbf{V}^{*} - \mathbf{X}\mathbf{W}\|
\end{aligned} \tag{A.2}$$

where, $\mathbf{X}^{*} = \mathbf{X}\mathbf{A}$ and $\mathbf{V}^{*} = \mathbf{A}^{\mathrm{T}}\mathbf{V}$. $\mathbf{P}\mathbf{L}^{1/2}$ is an known orthogonal matrix whose norm is a constant, and $\mathbf{L}^{1/2}\mathbf{P}^{\mathrm{T}}\mathbf{V}$ is a unit orthogonal matrix whose norm is also a constant that is not influenced by $\mathbf{V}$. Therefore, Eq. (A.2) holds.

**Appendix B**

*Proof*

Concerning the following optimal problem,

$$\mathbf{w}_j = \arg\min \|\mathbf{X}\mathbf{v}_j^* - \mathbf{X}\mathbf{w}_j\|^2 + \lambda \mathbf{w}^T \dot{\boldsymbol{\Omega}} \mathbf{w} + \lambda_{1,j} \|\mathbf{w}_j\|_1$$
$$s.t. \quad \mathbf{v}_j^{*T} \mathbf{v}_j^* = 1$$
(B.1)

We define an artificial data set $(\hat{\mathbf{y}}, \widehat{\mathbf{X}})$ by

$$\hat{\mathbf{y}} = \begin{pmatrix} \mathbf{X}^* \mathbf{v}_j^* \\ \mathbf{0} \end{pmatrix}, \quad \widehat{\mathbf{X}} = (1+\lambda)^{-1/2} \begin{pmatrix} \mathbf{X} \\ \lambda^{1/2} \mathbf{P} \mathbf{L}^{1/2} \mathbf{P}^T \end{pmatrix}$$
(B.2)

Let $\hat{\lambda} = \lambda_{1,j} / \sqrt{1+\lambda}$ and $\hat{\mathbf{w}}_j = \sqrt{1+\lambda}\, \mathbf{w}_j$. Then the above problem can be reformulated as

$$\mathbf{w}_j = \arg\min \|\hat{\mathbf{y}} - \widehat{\mathbf{X}}\hat{\mathbf{w}}\|^2 + \hat{\lambda} \|\hat{\mathbf{w}}_j\|$$
(B.3)

**Reference**

[1] Qin SJ. Process data analytics in the era of big data, *AIChE J.*, 2014, 60(9), 3092-3100.

[2] Efron B, Tibshirani R. Statistical analysis data analysis in the computer age, *Science*, 1991, 253(5018), 390-395.

[3] Zhao CH, Huang B. A Full-condition Monitoring Method for Nonstationary Dynamic Chemical Processes with Cointegration and Slow Feature Analysis, *AIChE J*, 2018, 64(5), 1662-1681.

[4] Sun H, Zhang SM, Zhao CH, Gao, FR. A sparse reconstruction strategy for online fault diagnosis in nonstationary processes with no a priori fault information. Ind. Eng. Chem. Res.,2017, 56(24), 6993-7008.

[5] Song B, Shi HB, Ma Y, Wang J. Multisubspace principal component analysis with local outlier factor for multimode process monitoring. Ind. Eng. Chem. Res., 2014, 53(42), 16453-16464.

[6] Nomikos P, MacGregor JF. Monitoring batch processes using multiway principal component analysis, *AIChE J.*, 1994, 40, 1361−1375.

[7] Cao Z, Zhang R, Lu J, Gao F, Online identification for batch processes in closed loop incorporating priori controller knowledge, Computers & Chemical Engineering, 2016, 90, 222-233.

[8] Cao Z, Yang Y, Yi H, Gao F. Priori knowledge-based online batch-to-batch identification in a closed loop and an application to injection molding, Ind. Eng. Chem. Res., 2016, 55(32), 8818-8829.

[9] Cao Z, Zhang R, Lu J, Gao F, Two-time dimensional recursive system identification incorporating priori pole and zero knowledge, Journal of Process Control, 2016, 39, 100-110.

[10] Zhao SY, Huang B, Liu F. Detection and diagnosis of multiple faults with uncertain

parameters. IEEE Trans. Contr. Syst. Tech., 2017, 25(5), 1873-1881.

[11] Zhao CH, Gao FR. Fault subspace selection approach combined with analysis of relative changes for reconstruction modeling and multifault diagnosis, *IEEE Trans. Contr. Syst. Technol.*, 2016, 24(3), 928-939.

[12] Yu WK, Zhao CH, Sparse exponential discriminant analysis and its application to fault diagnosis, *IEEE Trans. Ind. Electron.*, 2018, 65(7), 5931-5940.

[13] Qin Y, Zhao CH, Gao FR. An Iterative two-step sequential phase partition (ITSPP) method for batch process analysis and online monitoring, *AIChE J*, 2016, 62(7), 2358-2373.

[14] Westerhuis JA, Kourti T, MacGregor JF. Analysis of multiblock and hierarchical PCA and PLS models, *J. Chemom.* 1998, 12, 301−321.

[15] Zhang Y, Zhou H, Qin SJ, Chai TY. Decentralized fault diagnosis of large-scale processes using multiblock kernel partial least squares, *IEEE Trans. Ind. Inform*. 2010, 6(1), 3-10.

[16] Qin SJ, Valle S, Piovoso MJ. On unifying multiblock analysis with application to decentralized process monitoring, *J. Chemom.* 2001, 15, 715−742.

[17] Tong C, Song Y, Yan X. Distributed statistical process monitoring based on four-subspace construction and Bayesian inference, *Ind. &Eng. Chem. Res.*, 2013, 52, 5, 9897–9907.

[18] Jiang QC, Yan XF, Huang B. Performance-driven distributed PCA process monitoring based on fault-relevant variable selection and Bayesian inference, *IEEE Trans. Ind. Electron.* 2016, 63(1), 377-386

[19] Li WQ, Zhao CH, Gao FR. Linearity evaluation and variable subset partition based hierarchical process modeling and monitoring. *IEEE Trans. Ind. Electron.* 2018, 65(3), 2683-2692.

[20] Zhao CH, Gao FR. Fault-relevant Principal Component Analysis (FPCA) method for multivariate statistical modeling and process monitoring. *Chemom. Intell. Lab. Syst.*,

2014, 133, 1-16.

[21] Lee G, Han C, Yoon E, Multiple-fault diagnosis of the Tennessee Eastman process based on system decomposition and dynamic PLS, *Ind. & End. Chem. Res.* 2004, 43(25), 8037-8048.

[22] Chen J, Liu KC. Online batch process monitoring using dynamic PCA and dynamic PLS models. *Chem. Eng. Sci.*, 2002, 57, 63-75.

[23] Russell EL, Chiang LH, Braatz RD. Fault detection in industrial processes using canonical variate analysis and dynamic principal component analysis. *Chemmo. Intell. Lab. Syst.*, 2000, 41, 81-93.

[24] Ruiz-Carcel C, Cao Y, Mba D, Lao L, Samel RT. Statistical process monitoring of a multiphase flow facility. *Contr. Eng. Pract.* 2015, 42, 74-88.

[25] Shang C, Yang F, Gao X, et al. Concurrent monitoring of operating condition deviations and process dynamics anomalies with slow feature analysis, *AIChE J.* 2015, 61(11), 3666-3682.

[26] Shang C, Huang B, Yang F, Huang DX. Slow feature analysis for monitoring and diagnosis of control performance, *J. Process Control*, 2016, 39, 21-34.

[27] Bohmer W, Grunewalder S, Nickisch H, et al. Generating feature spaces for linear algorithms with regularized kernel slow feature analysis, *Mach. Learn*. 2012, 89, 67-86.

[28] Zou H, Hastie T, and Tibshirani R. Sparse principal component analysis. Journal of computational and graphical statistics, 2006, 15(2), 265-286.

[29] Efron B, Hastie T, Johnstone I, et al. Least angle regression. *The Annals of statistics*, 2004, 32(2), 407-499.

[30] Sprent P, Smeeton NC. Applied nonparametric statistical methods. CRC Press, 2016.

[31] Chen Q, Wynne RJ, Goulding P, et al. The application of principal component analysis and kernel density estimation to enhance process monitoring. *Contr. Eng. Pract*. 2000,

8(5), 531-543.

[32] Yin S, Ding SX, Haghani A, Hao H, Zhang P. A comparison study of basic data-driven fault diagnosis and process monitoring methods on the benchmark Tennessee Eastman process. *J. Process Control*, 2012, 22, 1567–1581.

[33] Kong XB, Liu XJ, Lee Kwang Y, An Effective nonlinear multivariable HMPC for USC power plant incorporating NFN-based modeling. *IEEE Trans. Ind. Electron.*, 2016, 12 (2), 555-566.

**List of Figure Captions**

Figure 1 Flow chart of the SSFA based iterative variable subset partition algorithm

Figure 2 Illustration of the proposed two-level based distributed modeling strategy

Figure 3 Slowness extents of SFs by SFA and SSFA for TE benchmark data

Figure 4 Slowness comparison results of different subsets for TE benchmark data based on (a) probability distributions of **SL** and (b) sign-rank test

Figure 5 Monitoring results of TE process for Fault #1 using (a) SFA (b) PCA (c) H-PCA-KPCA (d) the proposed algorithm

Figure 6 Monitoring results of TE process for Fault #2 using (a) SFA algorithm (b) PCA algorithm (c) H-PCA-KPCA algorithm and (d) the proposed algorithm

Figure 7 Coefficient along the first sparse loadings of different S&DL subsets

Figure 8 The first system slow features for (a) Case #1 and (b) Case #2 extracted by the proposed method in different S&DL subsets

Figure 9 Monitoring results of 1000 MW ultra-supercritical thermal power unit for (a) Case #1 and (b) Case #2 using the proposed method

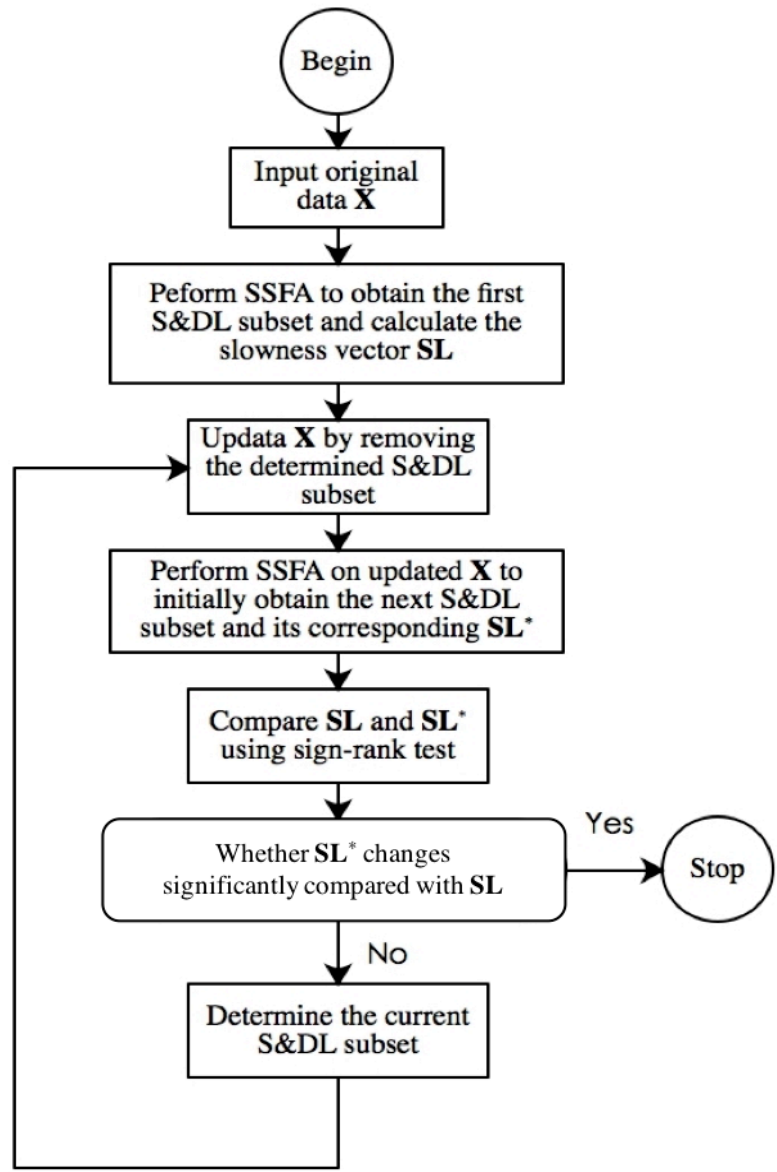

Figure 1 Flow chart of the SSFA based iterative variable subset partition algorithm

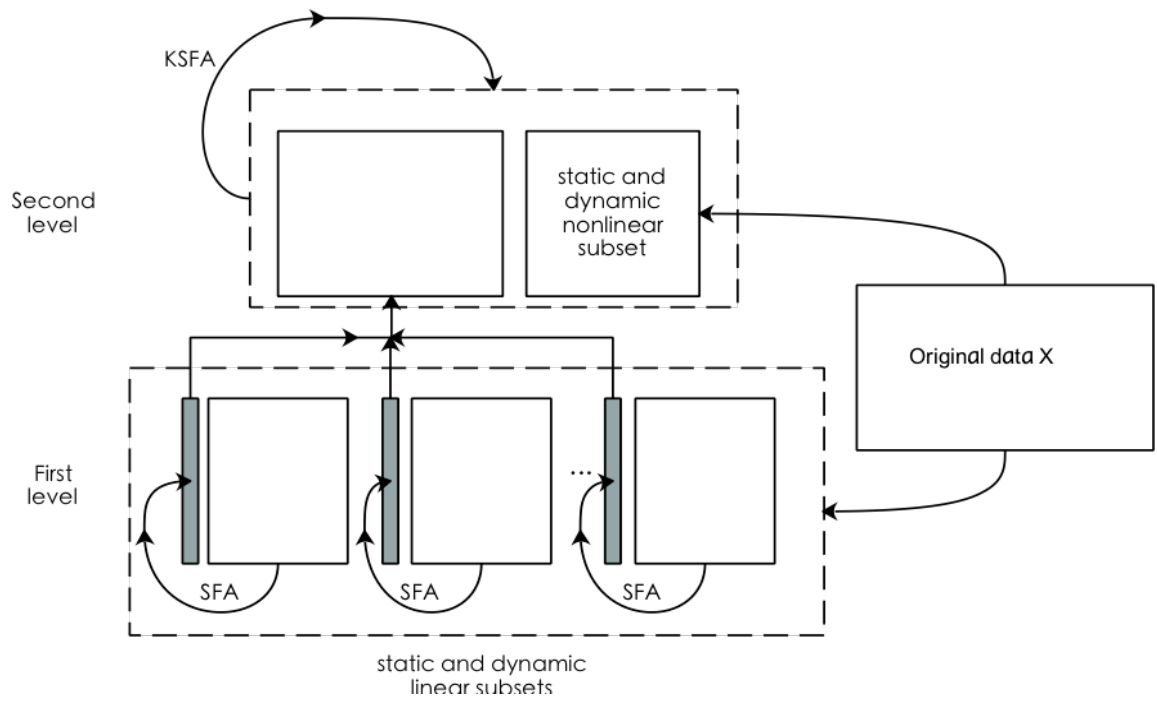

Figure 2 Illustration of the proposed two-level based distributed modeling strategy

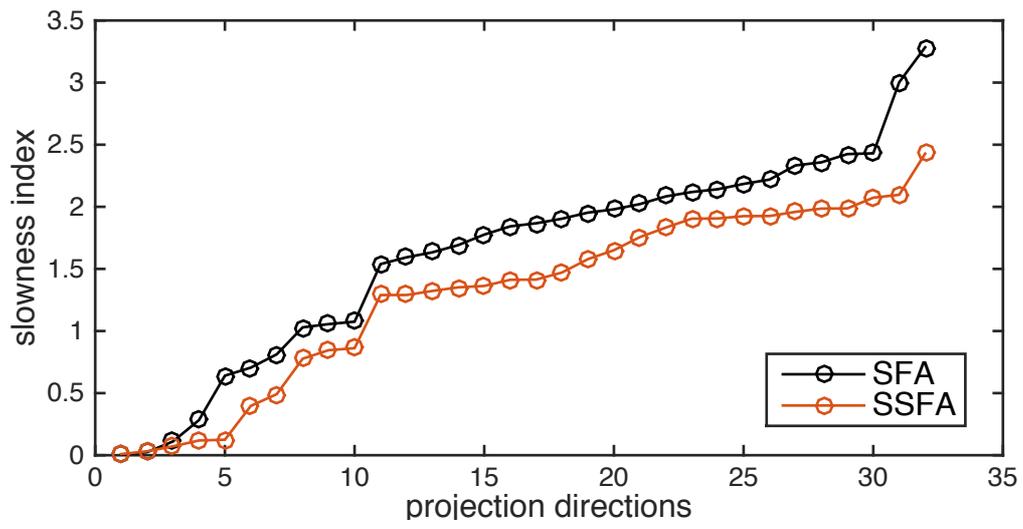

Figure 3 Slowness extents of SFs by SFA and SSFA for TE benchmark data

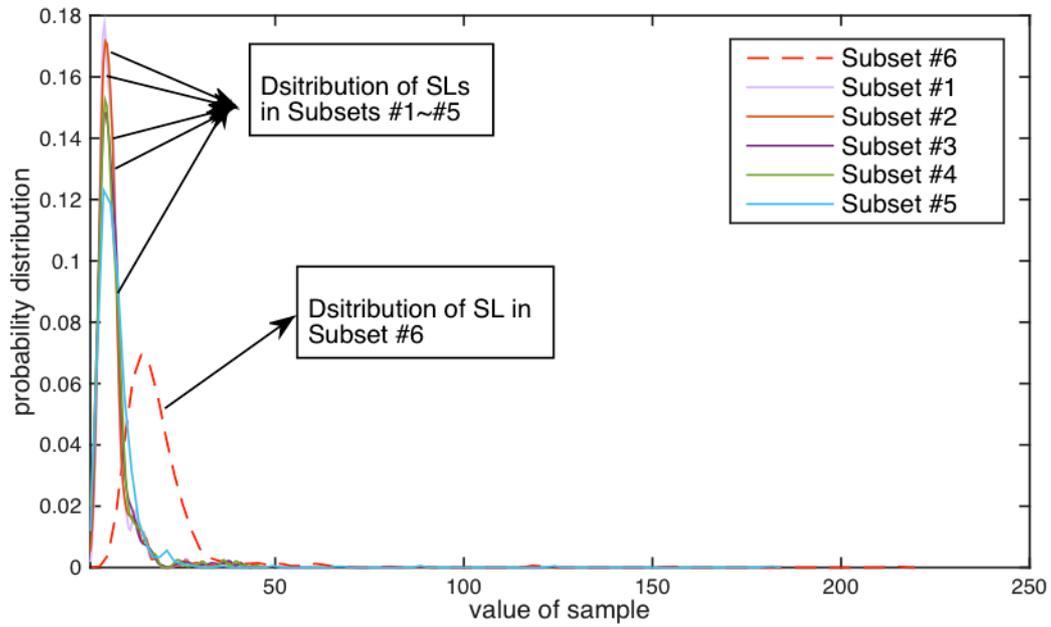

(a)

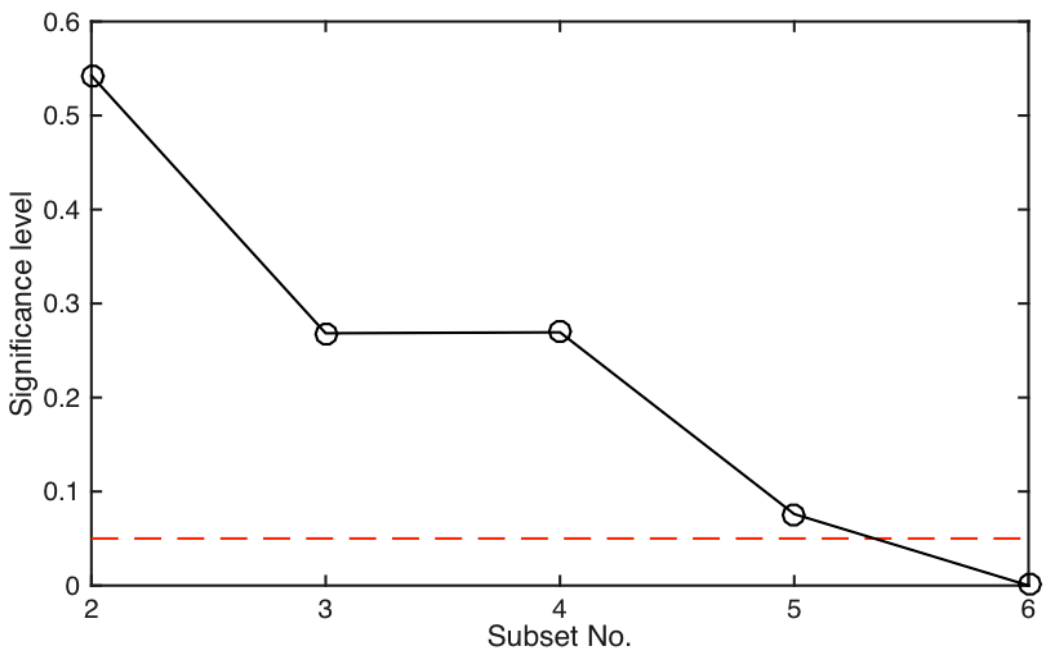

(b)

Figure 4 Slowness comparison results of different subsets for TE benchmark data based on

(a) probability distributions of **SL** and (b) sign-rank test

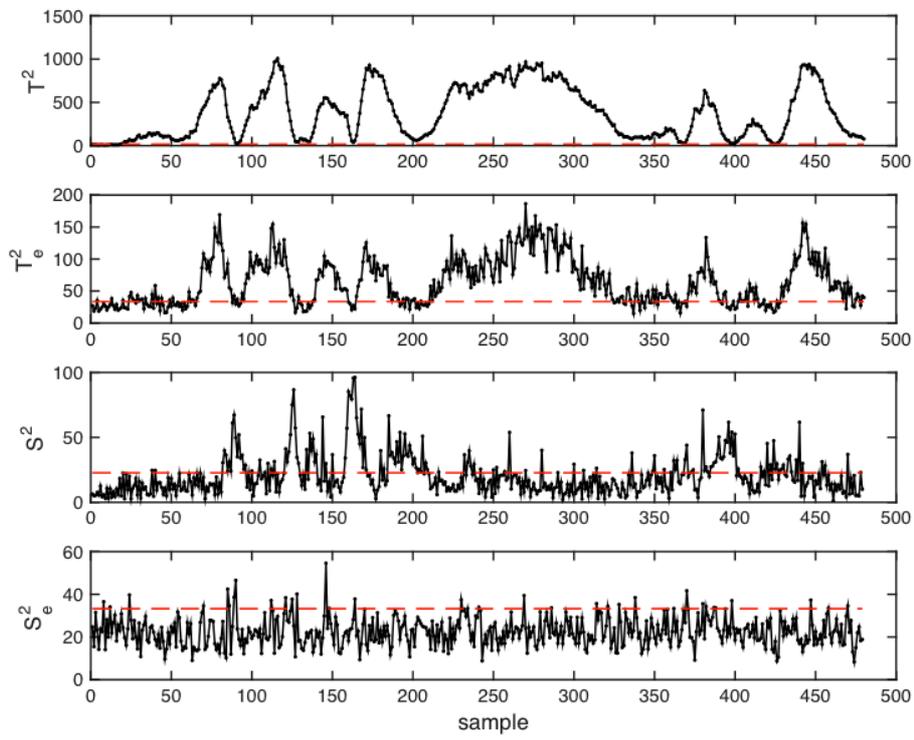

(a)

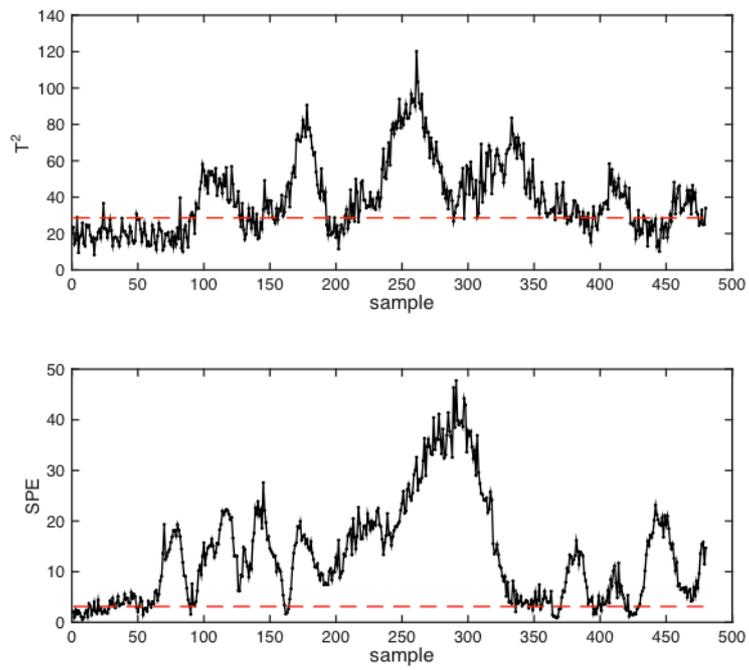

(b)

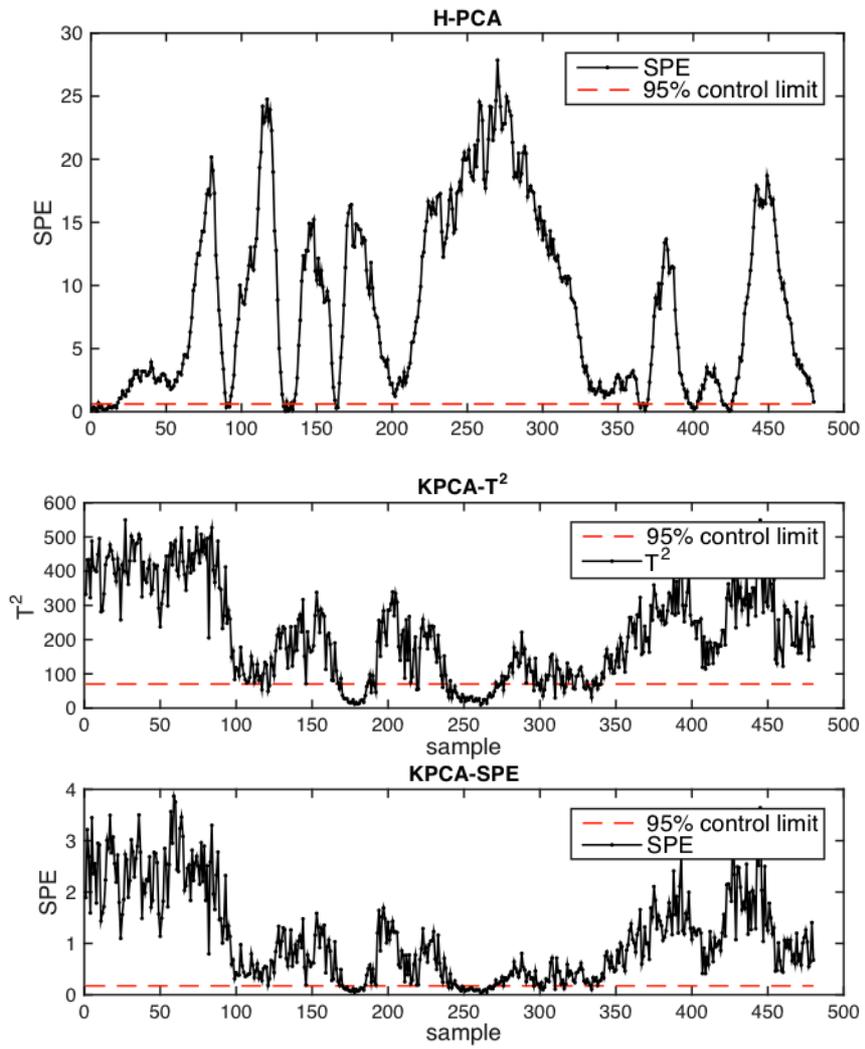

(c)

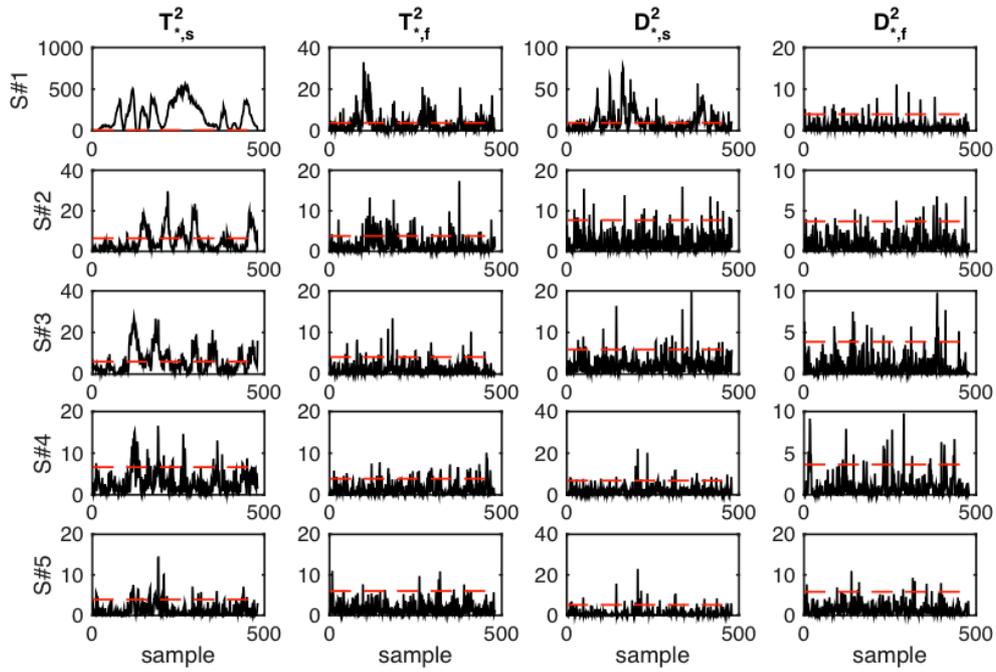

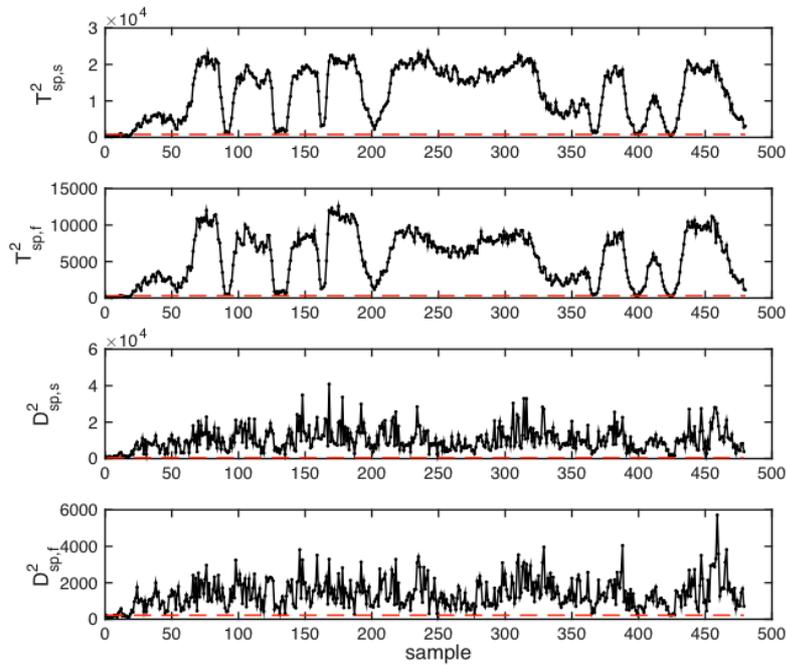

(d)

Figure 5 Monitoring results of TE process for Fault #1 using (a) SFA (b) PCA (c) H-PCA-KPCA (top subplot: lower level results; middle and bottom subplots: upper level results) (d) the proposed algorithm (top subplot: lower level results; bottom subplot: upper level results) (Red dashed line: 95% control limits; black dotted line: monitoring index)

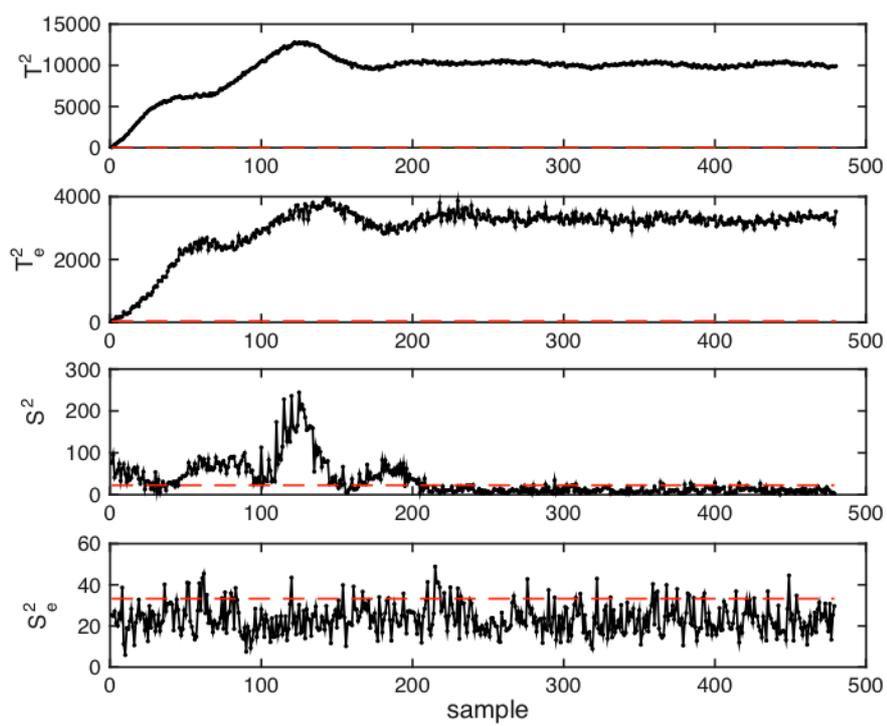

(a)

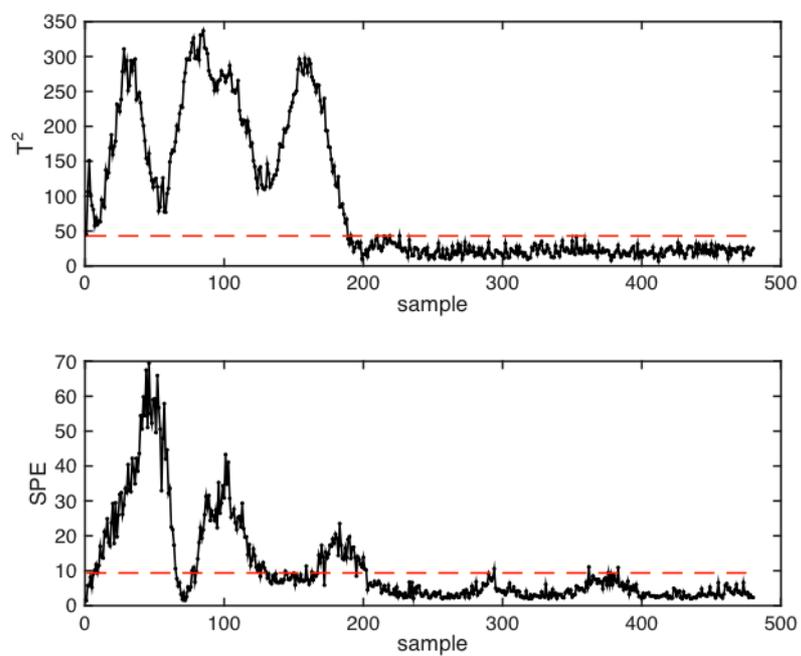

(b)

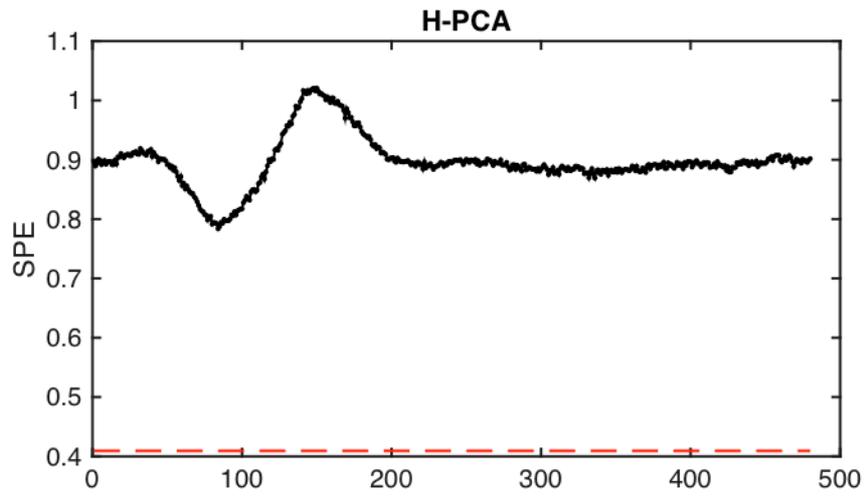

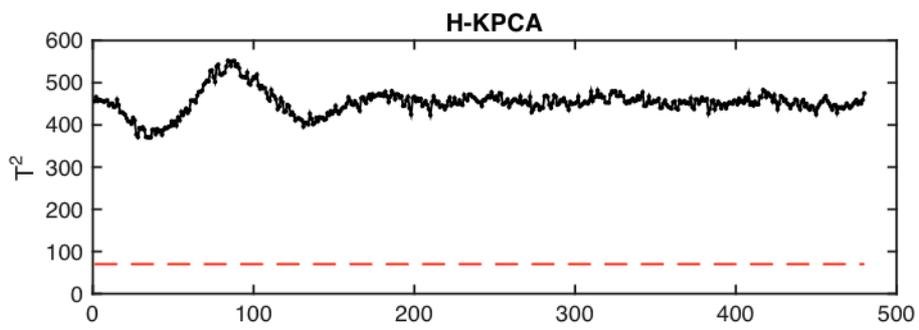

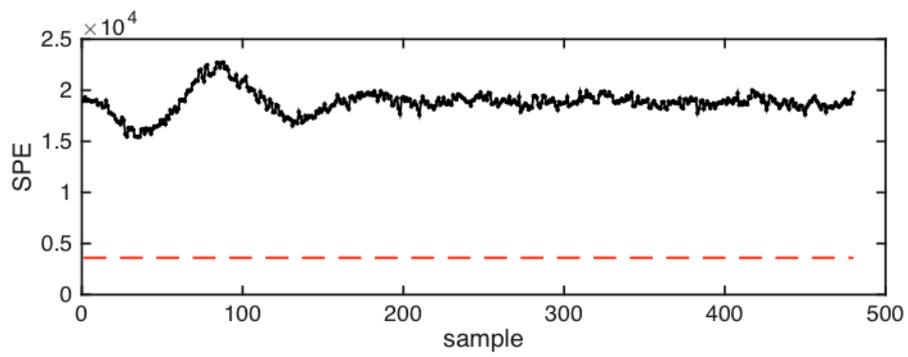

(c)

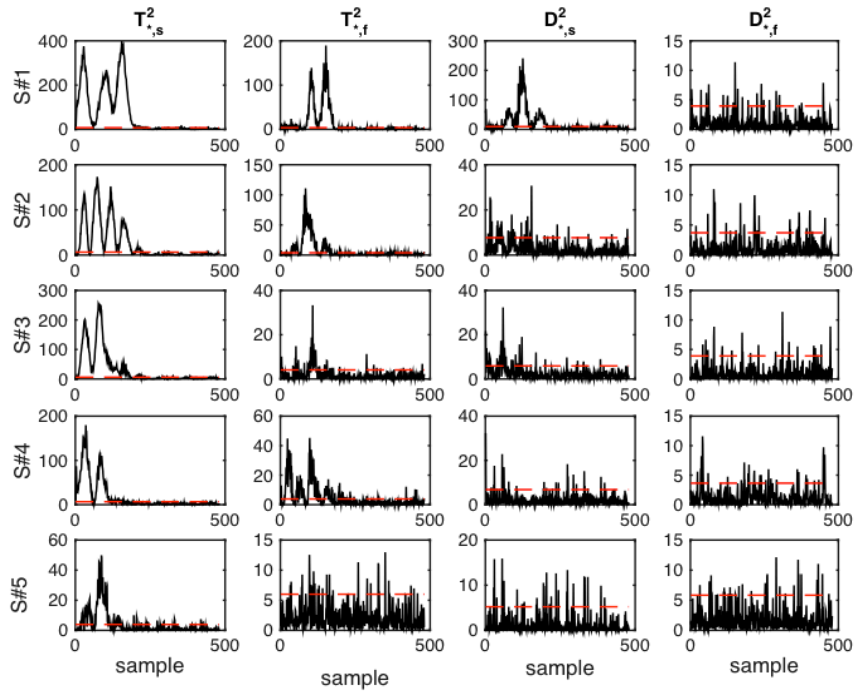

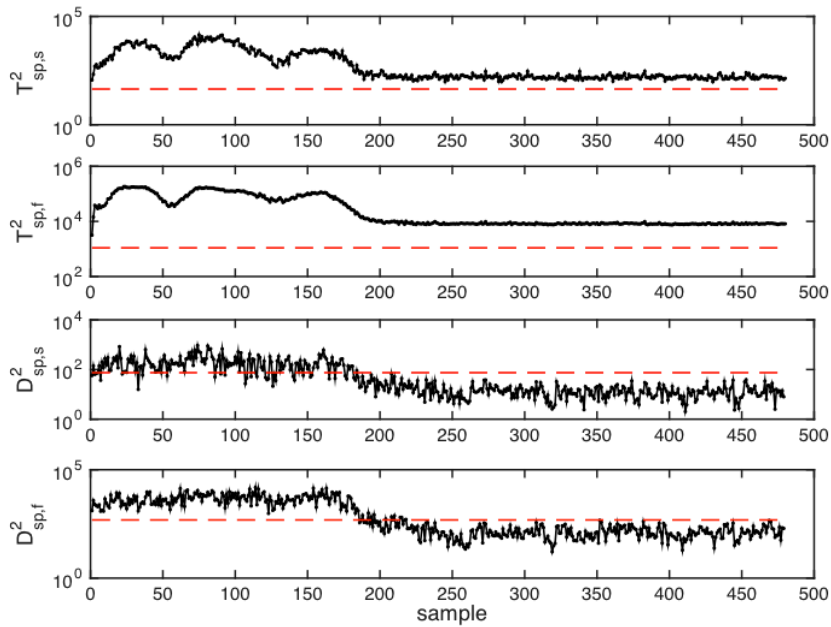

(d)

Figure 6 Monitoring results of TE process for Fault #2 using (a) SFA (b) PCA (c) H-PCA-KPCA (top subplot: lower level results; middle and bottom subplots: upper level results) (d) the proposed algorithm (top subplot: lower level results; bottom subplot: upper level results) (Red dashed line: 95% control limits; black dotted line: monitoring index)

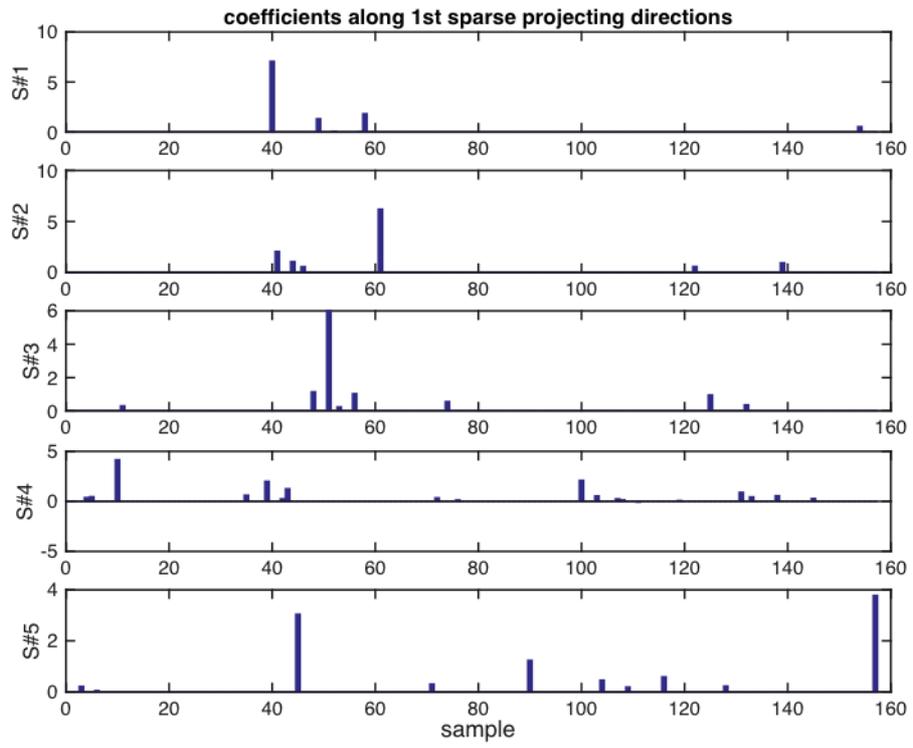

Figure 7 Coefficients along the first sparse loadings of different S&DL subsets

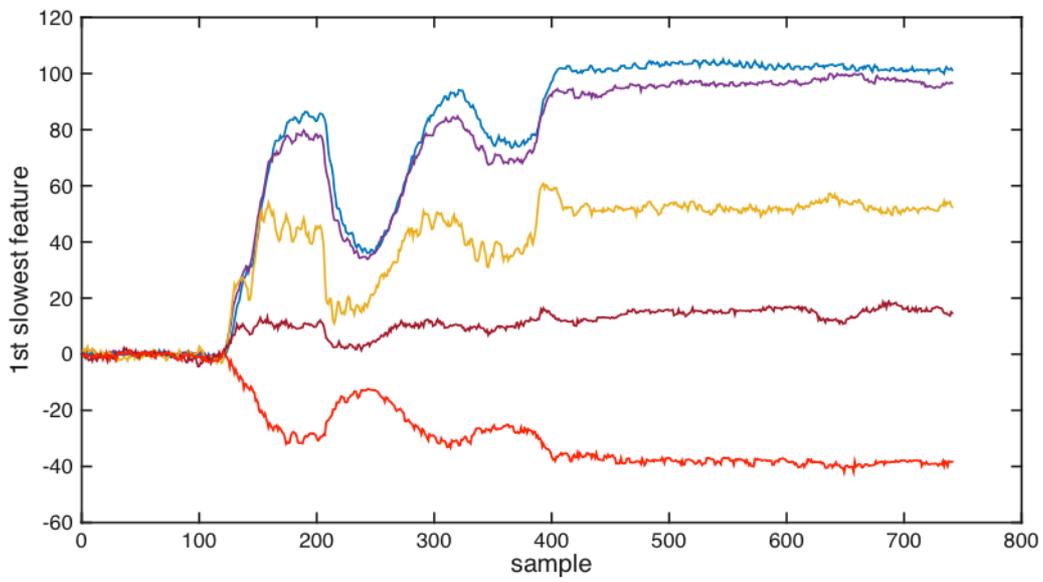

(a)

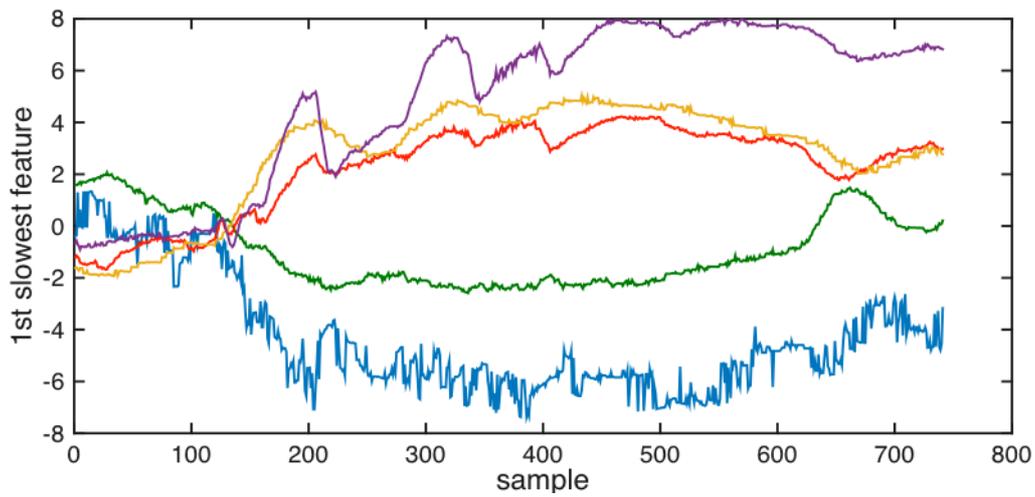

(b)

Figure 8 The first system slow features for (a) Case #1 and (b) Case #2 extracted by the proposed method in different S&DL subsets

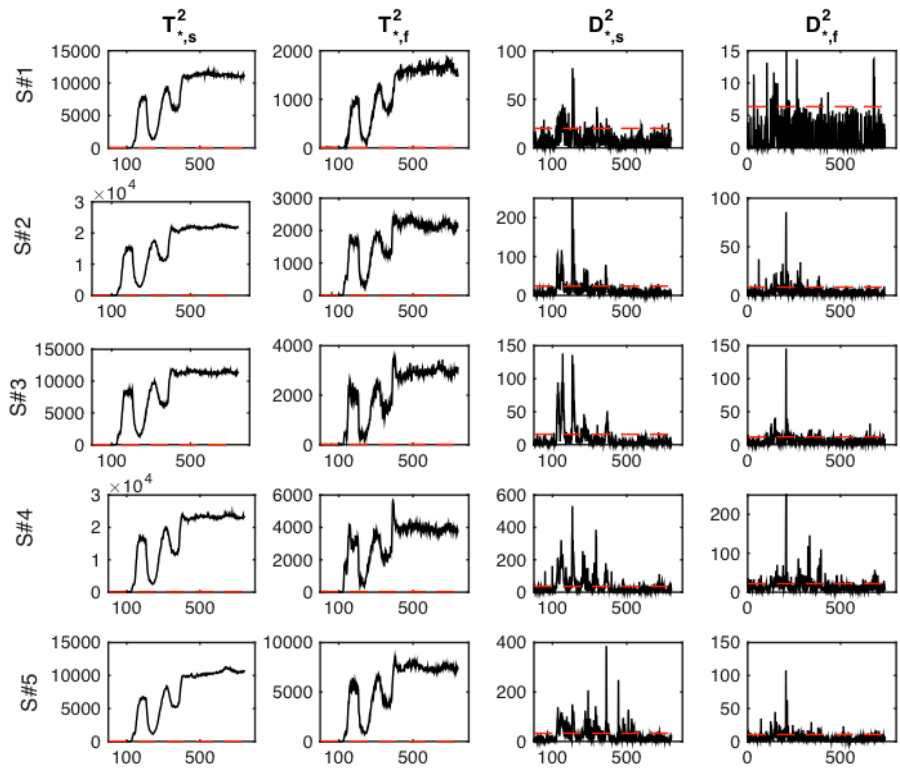

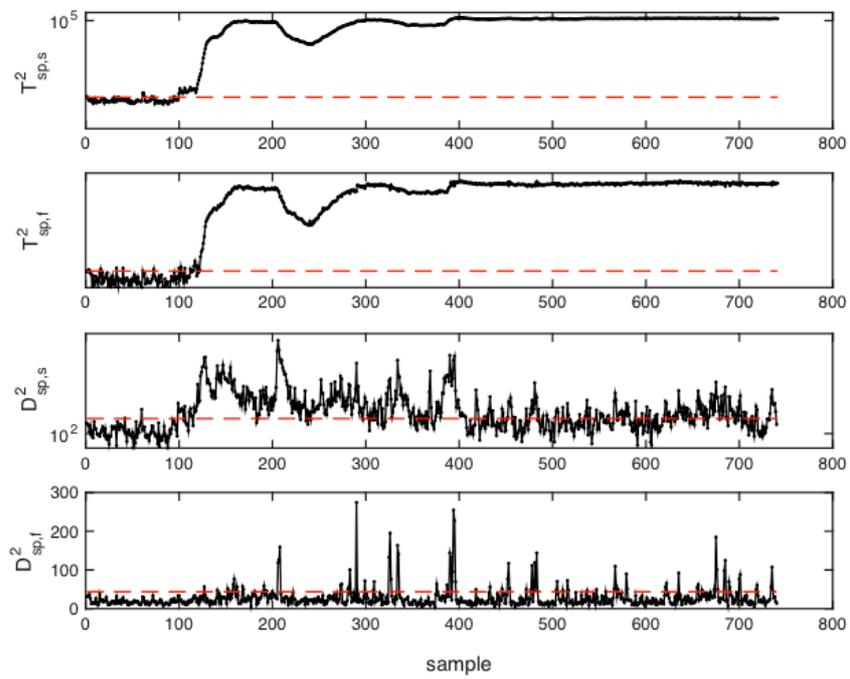

(a)

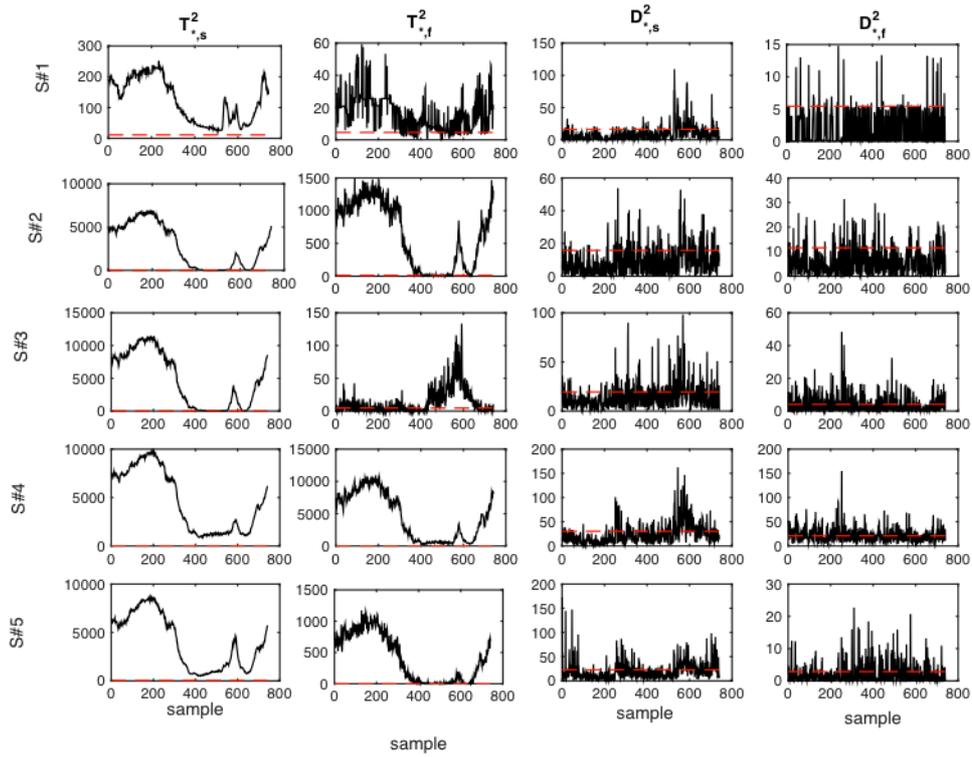

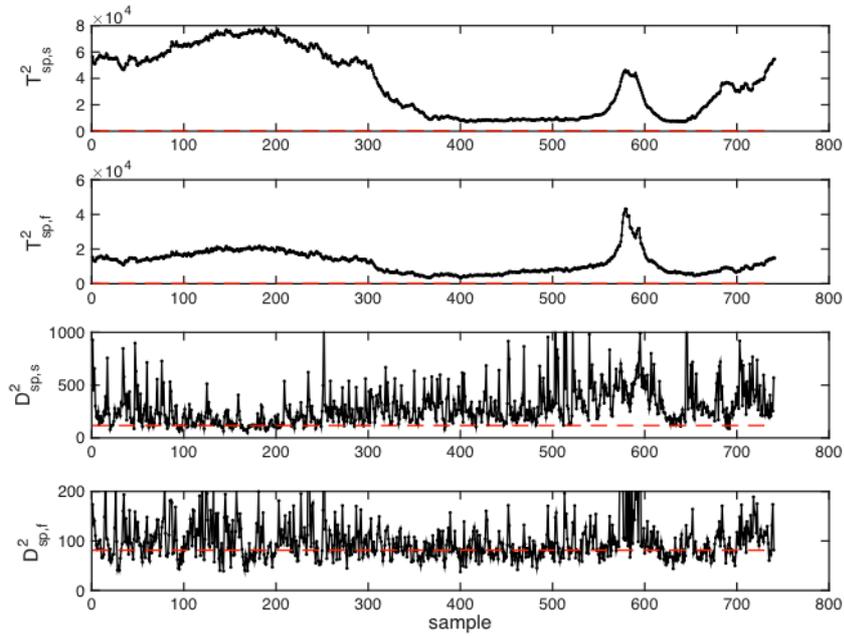

(b)

Figure 9 Monitoring results of 1000 MW ultra-supercritical thermal power unit for (a) Case #1 and (b) Case #2 using the proposed method (black dotted line: monitoring statistics; red dash line: 95% control limits)

Table I Description of different monitoring statistics

| Index | Description | |
|---|---|---|
| | First level (subset-wise) | Second level (plant-wide) |
| s | Local linear static system variations under certain operating conditions which are used to detect operating condition deviations in each subset | Global static nonlinear system variations under certain operating conditions which are used to detect plant-wide operating condition deviations |
| i | Local linear static process correlations under certain operating conditions which are used to check local operating condition deviations. | Global static nonlinear correlations under certain operating conditions which are used to detect plant-wide operation condition deviations. |
| c | Local linear dynamic variations of system information which are used to monitor control action in each subset | Global temporal variations of nonlinear system information which are used to monitor the overall control action of the whole process. |
| d | Local linear dynamic variations of noises which are used to monitor control action in each subset | Global temporal behaviors of nonlinear noises which are used to monitor the overall control action of the whole process. |

'$*$' is used to indicate the indexes of both the first level and second level

Table II TE process data: loadings of the first five SFs by SSFA

|  | SF#1 | SF#2 | SF#3 | SF#4 | SF#5 |
|---|---|---|---|---|---|
| Number of nonzero variables | 4 | 3 | 4 | 3 | 3 |
| Coefficients of nonzero variables | #1: 1.1681<br>#18: -6.6979<br>#19: -1.3970<br>#30: -2.4897 | #2: -0.9413<br>#16: -0.1189<br>#30: 5.5498 | #1: -0.6151<br>#16: -0.1142<br>#19: 3.6940<br>#24: -0.0279 | #7: 0.4174<br>#13: 0.2786<br>#16: 2.2374 | #20: -2.3698<br>#23: 0.0051<br>#26: -0.5211 |
| Slowness index | 0.0047 | 0.0317 | 0.0715 | 0.1187 | 0.1250 |

Table III Variable subset partition results for TE process

| S&DL Subset | Variable No. | S&DL Subset | Variable No. |
|---|---|---|---|
| #1 | 20,24,26 | #4 | 1,18,19,30 |
| #2 | 7, 13, 16 | #5 | 10, 11, 21 |
| #3 | 14, 22, 27 | | |

Table IV Variable subset partition results for 1000 MW ultra-supercritical thermal power unit

| Subset No. | Variable No. | Subset No. | Variable No. |
| --- | --- | --- | --- |
| #1 | 40,49,52,57,58,59,154 | #4 | 4,5,10,35,39,42, 43,72 |
| #2 | 41,44,46,47,55,60,61,122,139 | | 76,100,103,107,108,111,119, 131,133,138,145 |
| #3 | 11,48,51,53,56,74,125,132 | #5 | 3,6,45,71,90,104,109,116,128,157 |